\documentclass[preprint,twocolumn]{aastex62}
\usepackage{cprotect}
\hypersetup{linkcolor={blue},citecolor={blue}}
\received{\today}
\revised{\today}
\accepted{\today}
\submitjournal{ApJ}

\shorttitle{Bright optical transients with CSM-interaction}
\shortauthors{Suzuki, Moriya, \& Takiwaki}

\begin{document}
\title{A systematic study on the rise time--peak luminosity relation for bright optical transients powered by wind shock breakout}

\correspondingauthor{Akihiro Suzuki}
\email{akihiro.suzuki@nao.ac.jp}

\author[0000-0002-7043-6112]{Akihiro Suzuki}
\affil{Division of Science, National Astronomical Observatory of Japan, 2-21-1 Osawa, Mitaka, Tokyo 181-8588, Japan}

\author[0000-0003-1169-1954]{Takashi J. Moriya}
\affil{Division of Science, National Astronomical Observatory of Japan, 2-21-1 Osawa, Mitaka, Tokyo 181-8588, Japan}
\affiliation{Center for Computational Astrophysics, National Astronomical Observatory of Japan, 2-21-1 Osawa, Mitaka, Tokyo 181-8588, Japan}
\affiliation{School of Physics and Astronomy, Faculty of Science, Monash University, Clayton, Victoria 3800, Australia}

\author[0000-0003-0304-9283]{Tomoya Takiwaki}
\affil{Division of Science, National Astronomical Observatory of Japan, 2-21-1 Osawa, Mitaka, Tokyo 181-8588, Japan}
\affiliation{Center for Computational Astrophysics, National Astronomical Observatory of Japan, 2-21-1 Osawa, Mitaka, Tokyo 181-8588, Japan}



\begin{abstract}
A number of astrophyical transients originating from stellar explosions are powered by the collision of the ejected material with the circumstellar medium, which efficiently produces thermal radiation via shock dissipation. 
We investigate how such interaction-powered transients are distributed in the peak bolometric luminosity vs the rise time phase space. 
Taking the advantage of less time-consuming one-dimensional simulations with spherical symmetry, we calculated more than 500 models with different circumstellar mass and radius, ejecta mass and energy, and chemical compositions. 
The peak bolometric luminosity, the total radiated energy, and the rise time of the interaction-powered emission are measured for each simulated light curve. 
We consider how these characteristic quantities are determined as a function of the model parameters and discuss possible implications for the observed populations of (potential) interaction-powered transients, such as type IIn supernovae and fast blue optical transients. 
\end{abstract}



\section{Introduction}\label{sec:introduction}
Modern unbiased transient surveys have revealed the universe filled with transient phenomena with a wide variety of brightness and evolutionary timescales. 
Most bright optical transients associated with the death of massive stars were believed to be powered by radioactive nuclei \citep{1969ApJ...157..623C}, which are produced by the terminal explosion of a massive star, i.e., core-collapse supernovae (CCSNe). 
Although it is true for normal SNe, it has become clear that there are various optical transients likely powered by different energy sources, such as the accretion power, the rotational energy of the central compact object, and the interaction with the circumstellar matter (CSM). 
Among them, the CSM interaction has long been considered as a major energy source for luminous type IIn SNe \citep{1990MNRAS.244..269S,1997ARA&A..35..309F,2017hsn..book..403S,2017hsn..book..843B}, which show some narrow line features in their spectra and thus imply the presence of slowly moving materials ahead of the SN ejecta. 
The interaction-powered emission has also been paid great attention since the modern transient surveys, e.g., Palomar Transient Factory (PTF:  \citealt{2009PASP..121.1395L}), ASAS-SN\citep{2017PASP..129j4502K}, Pan-STARRS\citep{2016arXiv161205560C}, and Subaru Hyper Suprime-Cam Subaru Strategic Program \citep{2018PASJ...70S...1M,2018PASJ...70S...4A,2019PASJ...71...74Y}, have discovered optical transients with short rising and declining timescales \citep[e.g.,][]{2010ApJ...724.1396O,2014ApJ...794...23D,2016ApJ...819...35A,2016ApJ...819....5T,2018MNRAS.481..894P,2019ApJ...885...13T,2020arXiv200302669T}. 
In addition to ground-based telescopes, the recent advance in space-based monitoring surveys, such as the {\it Kepler} mission \citep{2010Sci...327..977B,2010ApJ...713L..79K,2014PASP..126..398H}, also realized the early detection and subsequent follow-up observations of intriguing optical transients potentially explained by the interaction-powered emission, e.g., KSN 2015K \citep{2018NatAs...2..307R}. 

In the SN ejecta-CSM collision, the forward and reverse shocks developing in the interface separating the two media are responsible for dissipating the ejecta kinetic energy and converting it to the internal energy of the shocked media. 
When the forward shock driven by the fast-moving ejecta emerges from the photosphere located within the CSM (the wind shock breakout), the dissipated energy starts leaking into the interstellar space as interaction-powered emission. 
The shock breakout from a dense wind-like CSM or an extended stellar envelope has been considered as a plausible mechanism to produce both rapidly evolving luminous transients \citep[e.g.,][]{2010ApJ...724.1396O} and luminous type IIn SNe \citep[e.g.,][]{2011ApJ...729L...6C}. 
A lot of analytic and numerical light curve modelings of the (potential) interaction-powered transients have been performed \citep[e.g.,][]{2011MNRAS.415..199M,2013MNRAS.428.1020M,2012ApJ...759..108S,2012ApJ...746..121C,2012ApJ...757..178G,2019ApJ...884...87T,2019arXiv191208486T}. 
The physical mechanism responsible for such a massive mass ejection or an extended envelope in its final evolutionary stage is still unknown.  
Among various scenarios for the mass ejection, an energy deposition into the stellar envelope may be a key and is paid a lot of attention so far \citep{2010MNRAS.405.2113D,2012MNRAS.423L..92Q,2014ApJ...780...96S,2017MNRAS.470.1642F,2018MNRAS.476.1853F,2019ApJ...877...92O,2019MNRAS.485..988O,2019arXiv191209738K}. 
The emission properties of transients powered by the wind shock breakout should be related to the properties of the physical mechanism responsible to the production of the CSM. 

Some past and ongoing supernova surveys have accumulated statistical samples of interaction-powered SNe. 
\cite{2014ApJ...788..154O} have compiled 15 type IIn SNe from the PTF/iPTF. 
They claimed a possible correlation between the peak luminosity and the rising timescale. 
Recently \cite{2019arXiv190605812N} have compiled an untargeted type IIn SN sample from the PTF/iPTF, which consists of 42 SNe. 
Their statistical analysis clarified a correlation between the rising timescale and the declining rate of the luminosity. 
Although luminous type IIn SNe are generally long-lasting, they claim that the correlation between the rising timescale and the peak luminosity is weak. 
On-going high-cadence optical transient surveys, such as Zwicky Transient Facility (ZTF; \citealt{2019PASP..131a8002B}), will also increase the sample size in the near future. 

The future deployment of Large Synoptic Survey Telescope (LSST, also known as Vera C. Rubin Observatory)\footnote{https://www.lsst.org} will further boost the potential of detecting interaction-powered transients even at high redshifts. 
In the coming LSST era, however, it would not be practical to conduct multiple follow-up spectroscopic observations for all the transients of interest discovered by the survey. 
Instead, some characteristic quantities from multi-band photometric observations, such as the peak luminosity, the rising and declining timescales, and the color evolution, would be provided for a lot of potentially intriguing transients. 
Therefore, it is beneficial to investigate how various populations of optical transients distribute in the phase space of characteristic quantities. 

As for type IIn SNe, \cite{2014ApJ...789..104O} first studied possible correlations in some light-curve properties based on the wind shock breakout scenario. 
\cite{2014ApJ...790L..16M} also used self-similar solutions for the ejecta-CSM interaction to obtain scaling relations for some light-curve properties. 
\cite{2017ApJ...849...70V} have used (semi-)analytic light curve models for various optical transients, including interaction-powered SNe, to investigate how the peak luminosity-duration phase space is filled with those transients. 
However, these studies are based on analytical models with some simplified treatments of the emission processes. 

In this work, we perform 1D radiation-hydrodynamic simulations of SN ejecta interacting with a spherical, wind-like CSM. 
Using more than 500 numerical models, we investigate how they behave in the peak luminosity vs rise time plot and what kinds of information regarding interaction-powered transients could be exploited from observed samples. 

This paper is organized as follows. 
In Section \ref{sec:method}, we describe the setups of our numerical simulations and introduce important free parameters characterizing the simulations. 
The simulation results are presented in Section \ref{sec:result}. 
We find that some analytic scaling relations are useful in understanding the numerical results. 
Section \ref{sec:phase_space} and Appendix \ref{sec:self_similar} provide the derivations of the analytic scaling relations. 
In Section \ref{sec:t_rise_vs_L_peak}, we consider the distribution of interacting transients in the peak luminosity vs duration phase space and their potential applications to current and future samples of interaction-powered transients.
Finally, we conclude this paper in Section \ref{sec:summary}.

\section{Numerical Setups}\label{sec:method}
Our numerical models are mostly based on our previous work \citep{2016ApJ...825...92S,2019ApJ...887..249S}. 
We have developed an Eulerian radiation-hydrodynamics code equipped with an adaptive mesh refinement technique and applied it to bipolar SN shock breakout \citep{2016ApJ...825...92S} and interacting SNe with spherical and disk-like CSMs \citep{2019ApJ...887..249S}. 
In this work, we use the 1D version of the numerical code to conduct a series of simulations of spherical SN ejecta colliding with a spherical wind-like CSM. 
The numerical setups are also similar to our previous work. 
In the following, we briefly describe our numerical setups. 

\subsection{SN ejecta}
Initially, the SN ejecta is assumed to be freely expanding. 
In other words, the radial velocity of a layer located at radius $r$ at time $t$ is given by $v=r/t$. 
The initial time of the simulations is set to $t=t_0=1000$ s. 
The outermost layer of the ejecta is adjacent to the inner edge of the CSM at $r=R_\mathrm{in}=4.0\times 10^{12}$ cm. 
Thus, the maximum ejecta velocity is given by $v_\mathrm{max}=R_\mathrm{in}/t_0=4.0\times 10^9$ cm s$^{-1}$. 
The radial density structure is expressed in the following way,

\begin{equation}
    \rho_\mathrm{ej}(r)=\frac{f_3M_\mathrm{ej}}{4\pi v_\mathrm{br}^3t_0^3}g(r/t_0),
\label{eq:rho_sn}
\end{equation}
with
\begin{equation}
    g(v)=\left\{
    \begin{array}{ccl}
    \left(\frac{v}{v_\mathrm{br}}\right)^{-\delta}&\mathrm{for}&v\leq v_\mathrm{br},\\
    \left(\frac{v}{v_\mathrm{br}}\right)^{-m}&\mathrm{for}&v_\mathrm{br}<v\leq v_\mathrm{max},\\
    0&\mathrm{for}&v_\mathrm{max}<v,\\
    \end{array}
    \right.
\end{equation}
and
\begin{equation}
    f_l=\frac{(m-l)(l-\delta)}{m-\delta-(l-\delta)(v_\mathrm{br}/v_\mathrm{max})^{m-l}},
\end{equation}
\citep{1989ApJ...341..867C,1999ApJ...510..379M}. 
The exponent $\delta$ characterizes the inner density gradient and is usually set to $\delta=0$--$2$ ($\delta=1$ in this study) so that the ejecta mass does not diverge. 
The outer part of the ejecta usually has a steeper density gradient, $m=7$--$12$ ($m=10$ in this study). 
The break velocity $v_\mathrm{br}$, at which the inner and outer parts of the ejecta are connected, gives the characteristic velocity of the ejecta. 
For a given set of the ejecta mass $M_\mathrm{ej}$ and the initial kinetic energy $E_\mathrm{sn}$, the break velocity is expressed as follows,
\begin{equation}
    v_\mathrm{br}=
    \left(\frac{2f_5E_\mathrm{sn}}{f_3M_\mathrm{ej}}\right)^{1/2}
    \simeq \left[\frac{2(m-5)(5-\delta)E_\mathrm{sn}}{(m-3)(3-\delta)M_\mathrm{ej}}\right]^{1/2}.
\end{equation}
The ejecta mass and energy, $E_\mathrm{sn}$ and $M_\mathrm{ej}$, are therefore important free parameters. 
For $(\delta,m)=(1,10)$, the break velocity is calculated to be
\begin{equation}
    v_\mathrm{br}=1.2\times10^9\mathrm{cm}\ \mathrm{s}^{-1}
    \left(\frac{M_\mathrm{ej}}{1M_\odot}\right)^{-1/2}
    \left(\frac{E_\mathrm{sn}}{10^{51}\mathrm{erg}\ \mathrm{s}^{-1}}\right)^{1/2}.
\end{equation}

\subsection{Spherical CSM}
We assume a spherical CSM outside the SN ejecta, $r>R_\mathrm{in}$. 
 We assume a power-law CSM with the density proportional to $r^{-q}$:
\begin{eqnarray}
    \rho_\mathrm{csm}(r)&=&\frac{pM_\mathrm{csm}}{4\pi R_\mathrm{csm}^3\Gamma((3-q)/p)}
    \nonumber\\
    &&\times\left(\frac{r}{R_\mathrm{csm}}\right)^{-q}\exp\left[-\left(\frac{r}{R_\mathrm{csm}}\right)^p\right],
    \label{eq:rho_csm}
\end{eqnarray}
where $\Gamma(x)$ is a gamma function and the exponents $p$ and $q$ are set to $(p,q)=(10,2)$, i.e, we consider steady wind-like CSMs. 
The exponential factor in this expression realizes a smooth cut-off around $r=R_\mathrm{csm}$. 
The mass $M_\mathrm{csm}$ and the radius $R_\mathrm{csm}$ determine the characteristic density of the CSM. 
We treat these two quantities as free parameters. 
Initially, the temperature of the CSM is set to $T_\mathrm{csm}=10^4$ K. 

A dilute medium corresponding to a normal stellar wind is assumed outside the dense CSM:
\begin{equation}
    \rho_\mathrm{out}(r)=A_\mathrm{out}r^{-2},
\end{equation}
where $A_\mathrm{out}=5\times 10^{11}$ g cm$^{-1}$. 
The coefficient $A_\mathrm{out}$ is set to so small that it has no significant impact on the propagation of radiation and shocks. 
The wind component extends to the outer boundary of the computational domain at $r=1.28\times 10^{17}$ cm. 
The total mass of this outer component in the computational domain is only $\simeq 4\times 10^{-4}M_\odot$, which is much smaller than the assumed ejecta and CSM masses. 
Then, the initial density structure is given by
\begin{equation}
    \rho(r)=\rho_\mathrm{ej}(r)+\rho_\mathrm{csm}(r)+\rho_\mathrm{out}(r). 
\end{equation}

\subsection{Numerical domain}
The simulation covers the radial coordinate $r\in[0,1.28\times 10^{17}]$ cm. 
We employ an adaptive mesh refinement (AMR) technique to cover the expanding ejecta. 
The base AMR grid with the lowest resolution is composed of $1024$ uniform numerical cells. 
The maximum AMR level is initially set to $14$. 
As the ejecta expand with time, the maximum AMR level is decreased one by one for the purpose of saving computational costs. 
The relative numerical resolution compared with the physical scale of the ejecta is guaranteed. 
This numerical prescription is proven to work well in our previous work \citep{2019ApJ...887..249S}. 
In particular, a radiative shock is known to form a narrow high-temperature layer and a density spike in the immediate downstream (see Figure \ref{fig:evolution_mej10} below), which are hard to resolve. 
As we have shown in \cite{2019ApJ...887..249S}, the density spike is typically covered by several numerical cells. 

\subsection{Radiative processes}
Our numerical code solves radiation hydrodynamic equations under a gray approximation and local thermodynamic equilibrium.  
We assume that free-free emission/absorption is the dominant radiative process creating/destructing photons. 
The free-free opacity is given by
\begin{equation}
\kappa_\mathrm{a}=3.7\times 10^{22}\chi_\mathrm{ion}(1+X_\mathrm{h})(X_\mathrm{h}+X_\mathrm{he})\rho T_\mathrm{g}^{-7/2}\ \mathrm{cm^2\ g^{-1}},
\end{equation}
(the local density $\rho$ and the gas temperature $T_\mathrm{g}$ are in cgs units; see, e.g., \citealt{1979rpa..book.....R}).
Here $X_\mathrm{h}$ and $X_\mathrm{he}$ represent the hydrogen and helium mass fractions. 
The factor $\chi_\mathrm{ion}$ describes the reduction of the opacity due to partial ionization (see below). 
The electron scattering opacity is defined as follows,
\begin{equation}
    \kappa_\mathrm{s}=0.2(1+X_\mathrm{h})\chi_\mathrm{ion},
    \label{eq:kappa_es}
\end{equation}
(e.g., \citealt{1979rpa..book.....R}) by using the same ionization parameter. 
At temperatures below $6000$--$7000$ K, hydrogen recombination reduces the free-free and electron scattering opacity. 
\cite{2019ApJ...879...20F} suggest that the ionization degree $\chi_\mathrm{ion}$ proportional to $T^{\beta}$ with $\beta=11$ mimics the recombination effect. 
We employ this prescription with a modification and assume the following ionization parameter,
\begin{equation}
\chi_\mathrm{ion}=\frac{1}{1+(T_\mathrm{g}/T_\mathrm{rec})^{-\beta}},
\end{equation} 
with $\beta=11$. 
The constant and the power-law parts are smoothly connected for the numerical convenience rather than the sudden change in the temperature gradient in \cite{2019ApJ...879...20F}. 

We mainly consider hydrogen-rich media with the hydrogen and helium mass fractions of $X_\mathrm{h}=0.73$ and $X_\mathrm{he}=0.25$ throughout the numerical domain. 
The recombination temperature is assumed to be $T_\mathrm{rec}=7000$ K. 
We also consider the effect of the reduced electron scattering opacity in hydrogen-free media with $X_\mathrm{h}=0.0$ and $X_\mathrm{he}=0.98$. 
In hydrogen-free models, the recombination temperature is set to $T_\mathrm{rec}=1.2\times 10^{4}$ K, reflecting a higher recombination temperature of helium. 
Although this treatment is a significantly simplified approximation for helium recombination, our focus is to investigate the effect of the reduced electron scattering in hydrogen-free media on the light curve properties rather than precisely implementing physical processes.

We keep track of the outgoing radiative flux $F_\mathrm{r}$ at $r=R_\mathrm{obs}=10^{17}$ cm. 
The bolometric luminosity of the emission going through the spherical boundary at $r=R_\mathrm{obs}$ is simply given by
\begin{equation}
    L_\mathrm{bol}(t)=4\pi R_\mathrm{obs}^2F_r(t,R_\mathrm{obs}). 
\end{equation}

\subsection{Model parameters}
Taking the advantage of less time-consuming 1D spherical simulations, we carry out simulations with various sets of the model parameters. 
The most important parameters are the ejecta mass and energy, $M_\mathrm{ej}$ and $E_\mathrm{sn}$, and the CSM mass and radius, $M_\mathrm{csm}$ and $R_\mathrm{csm}$.   
Among them, the CSM mass $M_\mathrm{csm}$ predominantly determines the evolutionary timescale of the interaction-powered emission as we shall see below and at the same time highly uncertain. 
Therefore, we treat models with different $M_\mathrm{csm}$ but with fixed other parameters as one series of simulations. 
In a single series of the simulations, the CSM mass is changed by more than two orders of magnitudes from $0.1M_\odot$ up to $50M_\odot$. 
The adopted mass grid is as follows:
 for $M_\mathrm{csm}\leq 1M_\odot$, the CSM mass is increased by $0.1M_\odot$, 
 for $1M_\odot<M_\mathrm{csm}\leq 10M_\odot$, it is increased by $1M_\odot$, and for $10M_\odot<M_\mathrm{csm}\leq 50M_\odot$, it is increased by $10M_\odot$. 
A single model series thus consists of $23$ models with different $M_\mathrm{csm}$. 

We vary other physical model parameters, $M_\mathrm{ej}$, $E_\mathrm{sn}$, and $R_\mathrm{csm}$ by a factor of 10 or 20. 
Table \ref{table:model_description} provides the series names and the corresponding free parameters. 
For the same parameter set, we assume ejecta with two different chemical abundances, hydrogen-rich ($X_\mathrm{h}=0.73$ and $X_\mathrm{he}=0.25$) and hydrogen-free ejecta ($X_\mathrm{h}=0$ and $X_\mathrm{he}=0.98$) to investigate the effect of reduced electron scattering opacity. 
In total, we simulate and analyze 22 model series with 506 models.

In Table \ref{table:Rph}, we provide the photospheric radii (divided by $R_\mathrm{csm}$) for models with different $M_\mathrm{csm}$ and $R_\mathrm{csm}$.  
They are calculated by assuming fully ionized hydrogen-rich CSMs. 
For a massive and compact CSM, the photosphere is located around the cut-off radius $r=R_\mathrm{csm}$. 
\begin{table}
\begin{center}
  \caption{Model descriptions}
\begin{tabular}{lrrr}
\hline\hline
Series&$M_\mathrm{ej}[M_\odot]$&$E_\mathrm{sn}[10^{51}\mathrm{erg}]$&$R_\mathrm{csm}[10^{15}\mathrm{cm}]$
\\
\hline
\verb|M1E1R5|&$1.0$&$1.0$&$5.0$\\
\verb|M2E1R5|&$2.0$&$1.0$&$5.0$\\
\verb|M5E1R5|&$5.0$&$1.0$&$5.0$\\
\verb|M10E1R5|&$10.0$&$1.0$&$5.0$\\
\verb|M1E01R5|&$1.0$&$0.1$&$5.0$\\
\verb|M1E02R5|&$1.0$&$0.2$&$5.0$\\
\verb|M1E05R5|&$1.0$&$0.5$&$5.0$\\
\verb|M1E2R5|&$1.0$&$2.0$&$5.0$\\
\verb|M1E1R1|&$1.0$&$1.0$&$1.0$\\
\verb|M1E1R2|&$1.0$&$1.0$&$2.0$\\
\verb|M1E1R10|&$1.0$&$1.0$&$10.0$\\
\hline\hline
\end{tabular}
\label{table:model_description}
\end{center}
\end{table}

\begin{table}
\begin{center}
  \caption{Photospheric radii for fully ionized hydrogen-rich CSMs}
\begin{tabular}{lrrrr}
\hline\hline
&\multicolumn{4}{c}{$R_\mathrm{ph}/R_\mathrm{csm}\ \mathrm{for}\ R_\mathrm{csm}/10^{15}\mathrm{cm}=$}
\\
$M_\mathrm{csm}[M_\odot]$&
$1$&
$2$&
$5$&
$10$
\\
\hline
$0.1$&$0.815$&$0.565$&$0.183$&$0.0536$\\
$0.2$&$0.894$&$0.707$&$0.306$&$0.101$\\
$0.3$&$0.930$&$0.774$&$0.395$&$0.144$\\
$0.4$&$0.952$&$0.815$&$0.461$&$0.183$\\
$0.5$&$0.968$&$0.843$&$0.514$&$0.218$\\
$0.6$&$0.980$&$0.864$&$0.555$&$0.250$\\
$0.7$&$0.989$&$0.880$&$0.590$&$0.279$\\
$0.8$&$0.997$&$0.894$&$0.619$&$0.306$\\
$0.9$&$1.00$&$0.905$&$0.643$&$0.331$\\
$1.0$&$1.01$&$0.914$&$0.664$&$0.354$\\
$2.0$&$1.04$&$0.968$&$0.783$&$0.514$\\
$3.0$&$1.06$&$0.994$&$0.838$&$0.605$\\
$4.0$&$1.07$&$1.01$&$0.871$&$0.664$\\
$5.0$&$1.08$&$1.02$&$0.894$&$0.706$\\
$6.0$&$1.09$&$1.03$&$0.911$&$0.738$\\
$7.0$&$1.09$&$1.04$&$0.924$&$0.763$\\
$8.0$&$1.10$&$1.04$&$0.935$&$0.783$\\
$9.0$&$1.10$&$1.05$&$0.944$&$0.800$\\
$10$&$1.10$&$1.05$&$0.952$&$0.815$\\
$20$&$1.12$&$1.08$&$0.997$&$0.894$\\
$30$&$1.13$&$1.09$&$1.02$&$0.930$\\
$40$&$1.14$&$1.10$&$1.03$&$0.952$\\
$50$&$1.15$&$1.11$&$1.04$&$0.968$\\
\hline\hline
\end{tabular}
\label{table:Rph}
\end{center}
\end{table}

\section{Numerical result}\label{sec:result}
In this section, we present our numerical results. 
We mainly focus on simulations with hydrogen-rich ejecta. 
\subsection{Dynamical evolution\label{sec:dynamical_evolution}}
\begin{figure}[htbp]
\begin{center}
\includegraphics[scale=0.55]{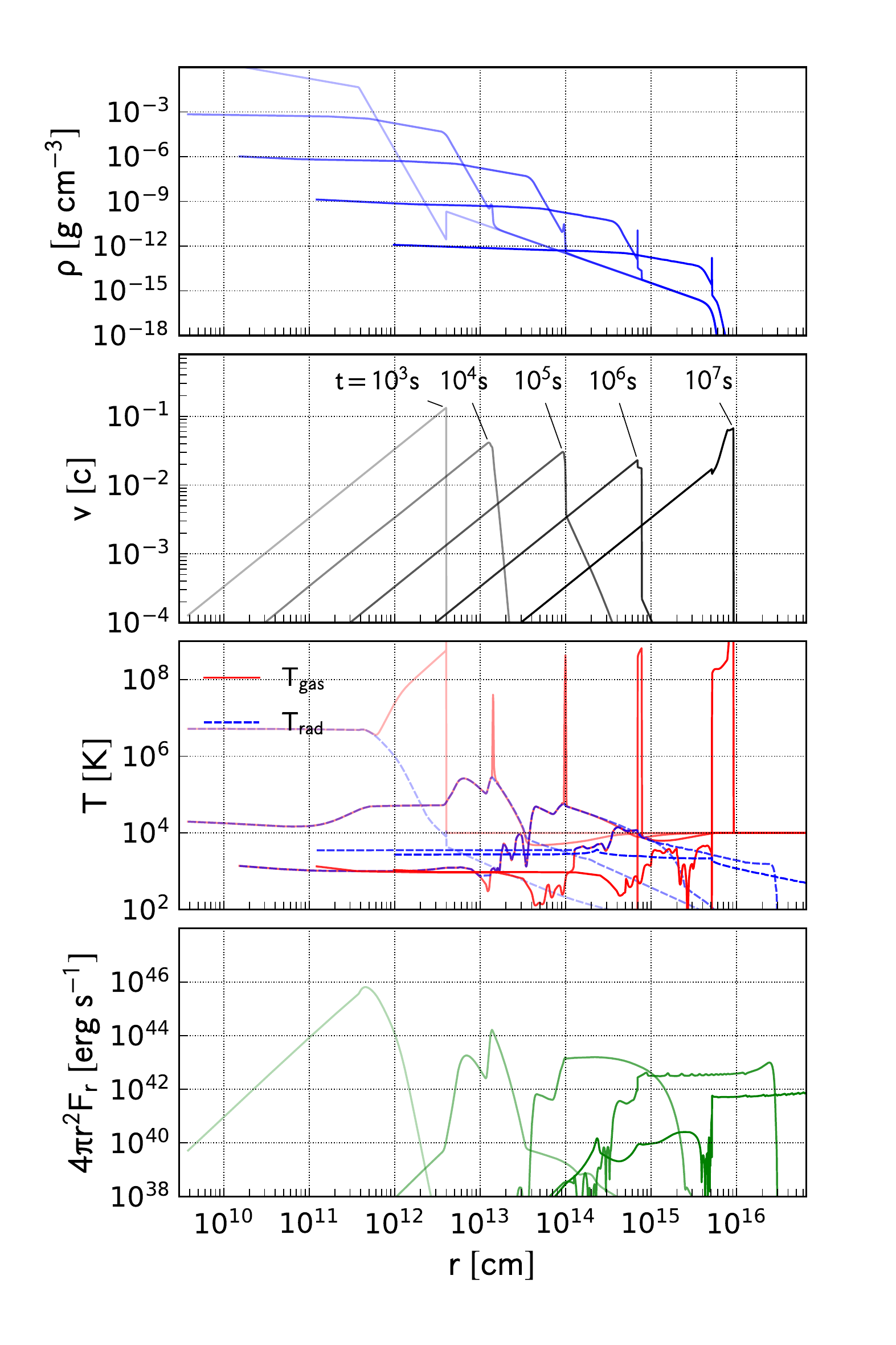}
\caption{Radial profiles of the density, the velocity, the gas and radiation temperatures, and the luminosity (from top to bottom) at $t=10^3$, $10^4$, $10^5$, $10^6$, and $10^7$ s. 
The result of the model with $M_\mathrm{ej}=10M_\odot$, $E_\mathrm{sn}=10^{51}$ erg, $M_\mathrm{csm}=0.1M_\odot$, and $R_\mathrm{csm}=5\times 10^{15}$ cm is presented. }
\label{fig:evolution_mej10}
\end{center}
\end{figure}
\begin{figure}[htbp]
\begin{center}
\includegraphics[scale=0.55]{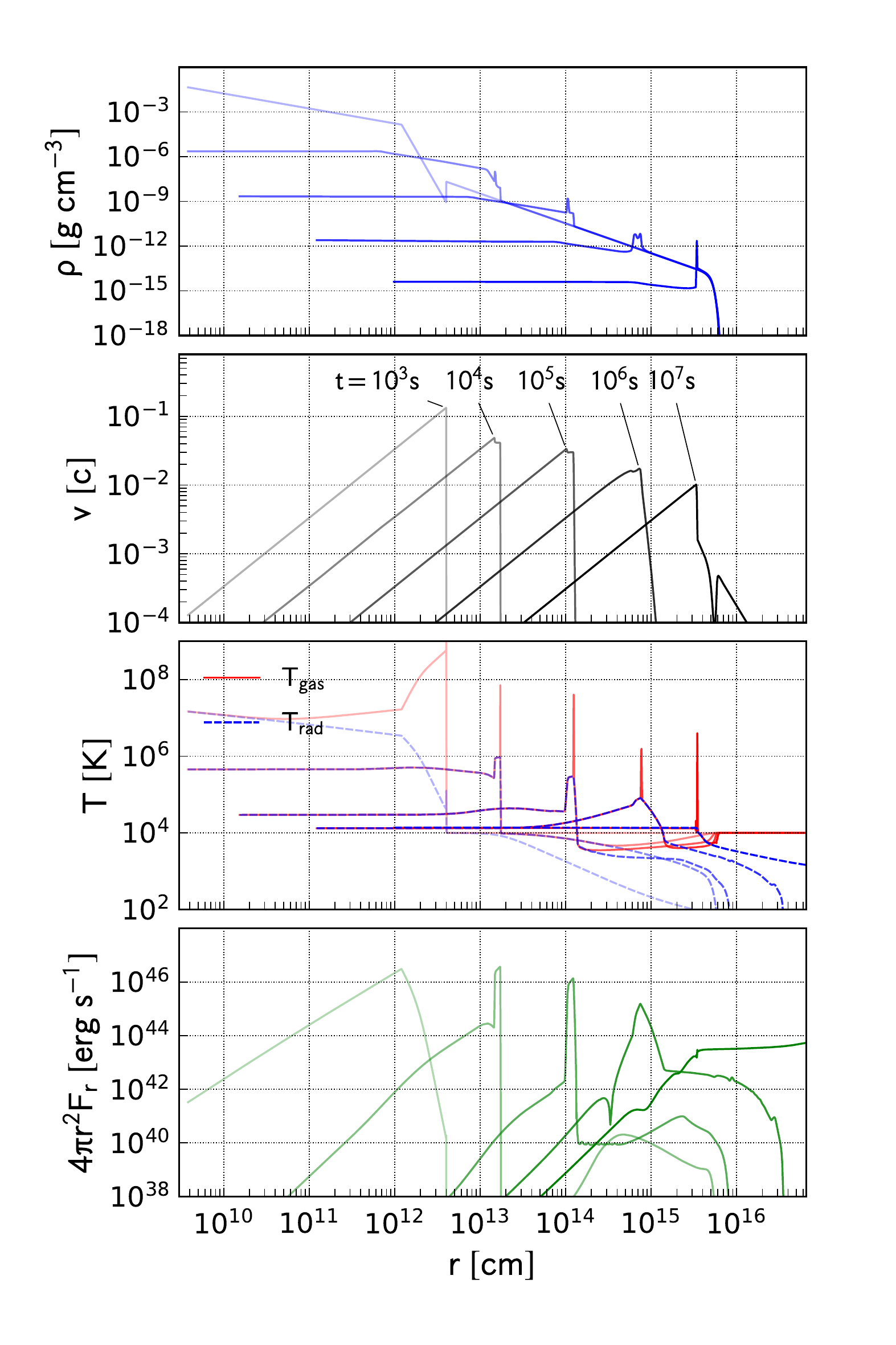}
\caption{Same as Figure \ref{fig:evolution_mej10}, but for the model with with $M_\mathrm{ej}=1M_\odot$, $E_\mathrm{sn}=10^{51}$ erg, $M_\mathrm{csm}=10M_\odot$, and $R_\mathrm{csm}=5\times 10^{15}$ cm.}
\label{fig:evolution_mej1}
\end{center}
\end{figure}

Figures \ref{fig:evolution_mej10} and \ref{fig:evolution_mej1} show how the radial distributions of some physical variables evolve with time for two cases with $M_\mathrm{ej}>M_\mathrm{csm}$ and $M_\mathrm{ej}<M_\mathrm{csm}$. 
In general, the radiative shock in the CSM evolves as follows. 
When the forward shock is still deeply embedded in the CSM, the radiation produced around the shock front is well confined in the post-shock region. 
Therefore, the post-shock gas behaves as an adiabatic gas with an effective adiabatic index of $4/3$. 
As the shock propagates in the CSM, however, it becomes easier for radiation to diffuse in the ambient gas due to the decreasing pre-shock density. 
Then, the shock starts suffering from radiative loss. 
The radiation front ahead of the shock front finally reaches the photosphere in the CSM, above which most photons can travel into the surrounding space without being absorbed nor scattered. 
This is the so-called shock breakout in the CSM and it happens when the photon diffusion velocity in the CSM exceeds the forward shock velocity. 
After the shock breakout, radiation in the post-shock region can easily escape through the photosphere. 
Even though photons can escape from the CSM, they experience multiple scattering episodes after their creation until they reach the photosphere. 
Therefore the photons are well thermalized and observed as thermal emission powered by the ejecta-CSM interaction. 
The thermalization efficiency is, in fact, sensitive to the local density of the CSM as we shall see below. 
The forward shock finally emerges from the outer edge of the CSM, at which the shock accelerates. 
Well after the emergence, the forward shock propagates in the dilute outer medium, where gas and radiation are only weakly coupled, and thus its efficiency to produce thermal photons suddenly drops. 
The temporal evolutions of the radial profiles shown in Figures \ref{fig:evolution_mej10} and \ref{fig:evolution_mej1} well reproduce the evolutionary stages described above. 

An important difference between the two models shown in Figure \ref{fig:evolution_mej10} and \ref{fig:evolution_mej1} is the CSM mass compared with the ejecta mass. 
In the former case, the CSM mass is much smaller than the ejecta mass, $M_\mathrm{csm}=0.1M_\mathrm{ej}$. 
Therefore, only a minor fraction of the ejecta is affected by the collision with the CSM up to $100$ days. 
As seen in Figure \ref{fig:evolution_mej10}, the reverse shock is still in the outer part of the ejecta at the time of the shock breakout ($t\simeq 10^5$s) and therefore most ejecta are still unshocked. 
In this case, the energy used for the interaction-powered emission is only a small fraction of the total kinetic energy of the SN ejecta. 
In Figure \ref{fig:evolution_mej1}, on the other hand, the massive CSM ($M_\mathrm{csm}=10M_\mathrm{ej}$) efficiently prevents the ejecta from expanding. 
As seen in the velocity profiles, the post-shock velocity significantly decreases from $v=0.1$c to $<0.01c$, while the forward shock is still in the CSM. 
In this case, the shock breakout occurs at several $10^6$ s. 
The most part of the ejecta has been swept up by the reverse shock until the shock breakout and thus almost all the initial kinetic energy of the ejecta has been dissipated and can be used as the radiation energy budget. 
This is because the CSM is much more massive than the SN ejecta. 
From these two different models, we expect that the characteristic properties of the interaction-powered emission exhibit different trends depending on $M_\mathrm{ej}>M_\mathrm{csm}$ or $M_\mathrm{ej}<M_\mathrm{csm}$.

\subsection{Light curves}
\begin{figure*}[tbp]
\begin{center}
\includegraphics[scale=0.5]{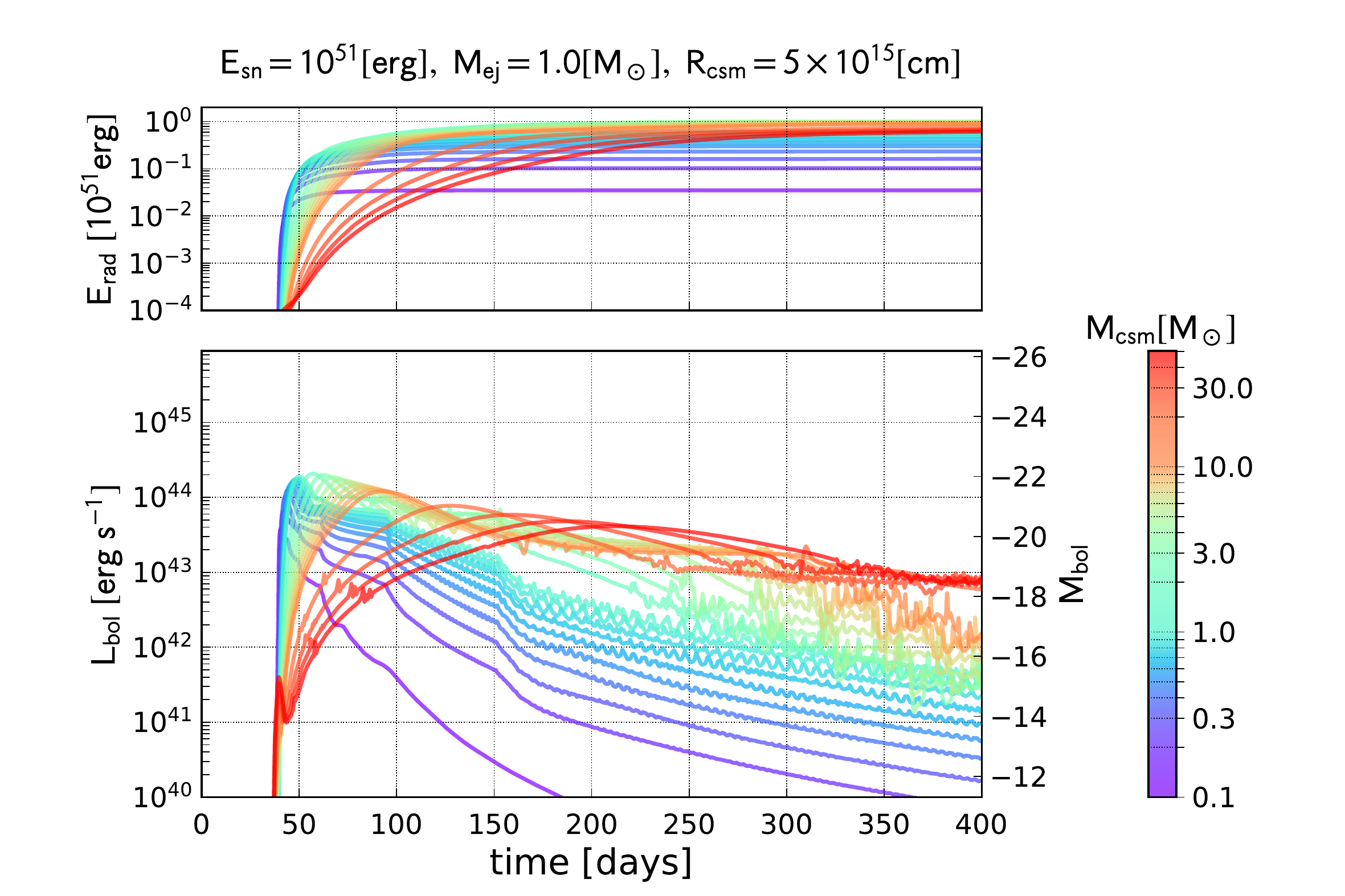}
\caption{Color-coded radiated energy (upper panel) and bolometric light curves (lower panel) of 23 models with $E_\mathrm{sn}=10^{51}$ erg, $M_\mathrm{ej}=1.0M_\odot$, and $R_\mathrm{csm}=5\times 10^{15}$ cm. 
The color of the curves represent the CSM mass. 
The CSM mass is increased from $M_\mathrm{csm}=0.1M_\odot$ to $M_\mathrm{csm}=50M_\odot$. }
\label{fig:lc}
\end{center}
\end{figure*}

\begin{figure*}
\begin{center}
\includegraphics[scale=0.5]{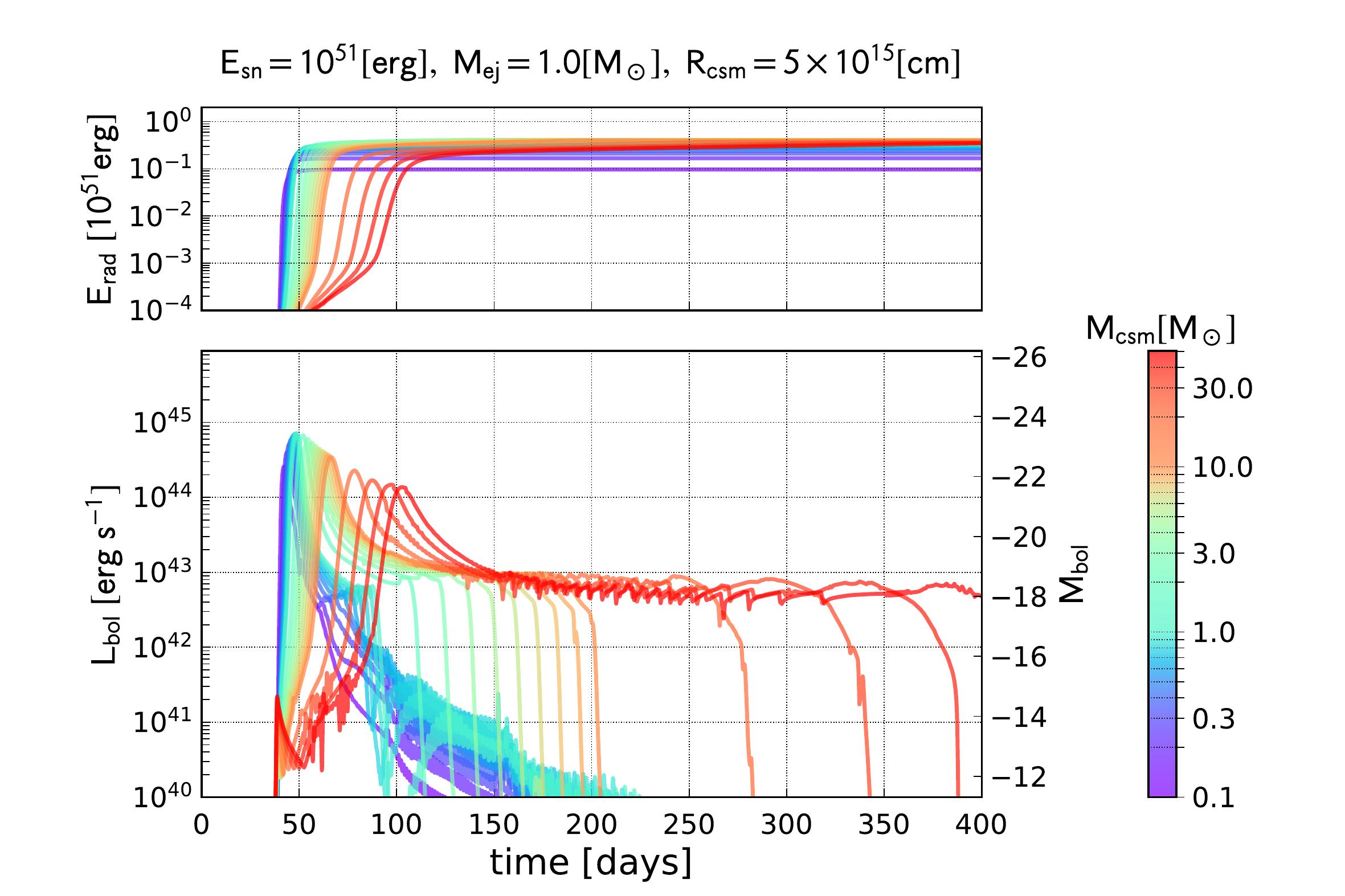}
\caption{Same as Figure \ref{fig:lc}, but for the models with a smaller $R_\mathrm{csm}=10^{15}$ cm. }
\label{fig:lc_r1e15}
\end{center}
\end{figure*}

Figures \ref{fig:lc} and \ref{fig:lc_r1e15} represent some example light curves for the two model series \verb|M1E1R5| and \verb|M1E1R1|. 
We also plot the cumulative radiated energy at time $t$:
\begin{equation}
    E_\mathrm{rad}(t)=\int_0^{t}L_\mathrm{bol}(t')dt'.
\end{equation}
As we have checked in our previous paper \citep{2019ApJ...887..249S}, our numerical simulations successfully reproduce the light curves of some type IIn SNe, e.g., SN 1998S, 2010jl, and 2006gy. 
The light curves are characterized by a smooth single peak, which separates each light curve into rising and declining parts. 
Although some light curves show artificial spikes in later epochs, which is produced by numerical treatments, we hereafter focus on the rising part, the peak bolometric luminosity, and the total radiated energy, and thus the numerical artifacts in the later declining phase less likely affect the results. 

As Figures \ref{fig:lc} and \ref{fig:lc_r1e15} demonstrate, models with different model parameters exhibit a wide variety of light curve properties. 
The peak bolometric luminosity ranges from a few $10^{43}$ erg s$^{-1}$ to $10^{45}$ erg s$^{-1}$. 
The evolutionary timescales are from a few to hundred days. 
In order to quantitatively analyze  these light curves, we introduce the following quantities: (1) the peak bolometric luminosity $L_\mathrm{bol,peak}$, (2) the total radiated energy $E_\mathrm{rad}$, and (3) the rise time $t_\mathrm{rise}$. 
The peak luminosity is defined as the maximum value of a given bolometric light curve. 
The radiated energy is obtained by integrating the bolometric light curve up to the end of the simulation at $t=6\times 10^7$ s. 
In order to determine the time $t_\mathrm{i}$ at which a light curve starts rising, we set a threshold luminosity $L_\mathrm{bol,th}=0.01L_\mathrm{bol,peak}$. 
We also define the peak time $t_\mathrm{peak}$ as the time of the maximum luminosity, $L_\mathrm{bol}(t_\mathrm{peak})=L_\mathrm{bol,peak}$. 
Then, $t_\mathrm{i}(<t_\mathrm{peak})$ is defined as the time closest to $t_\mathrm{peak}$ and  satisfying $L_\mathrm{bol}(t_\mathrm{i})=L_\mathrm{bol,th}$. 
Then, the rise time $t_\mathrm{rise}$ is given by the difference in the two epochs, $t_\mathrm{rise}=t_\mathrm{peak}-t_\mathrm{i}$.

\subsection{Dependence on CSM mass}
\begin{figure*}
\begin{center}
\includegraphics[scale=0.45]{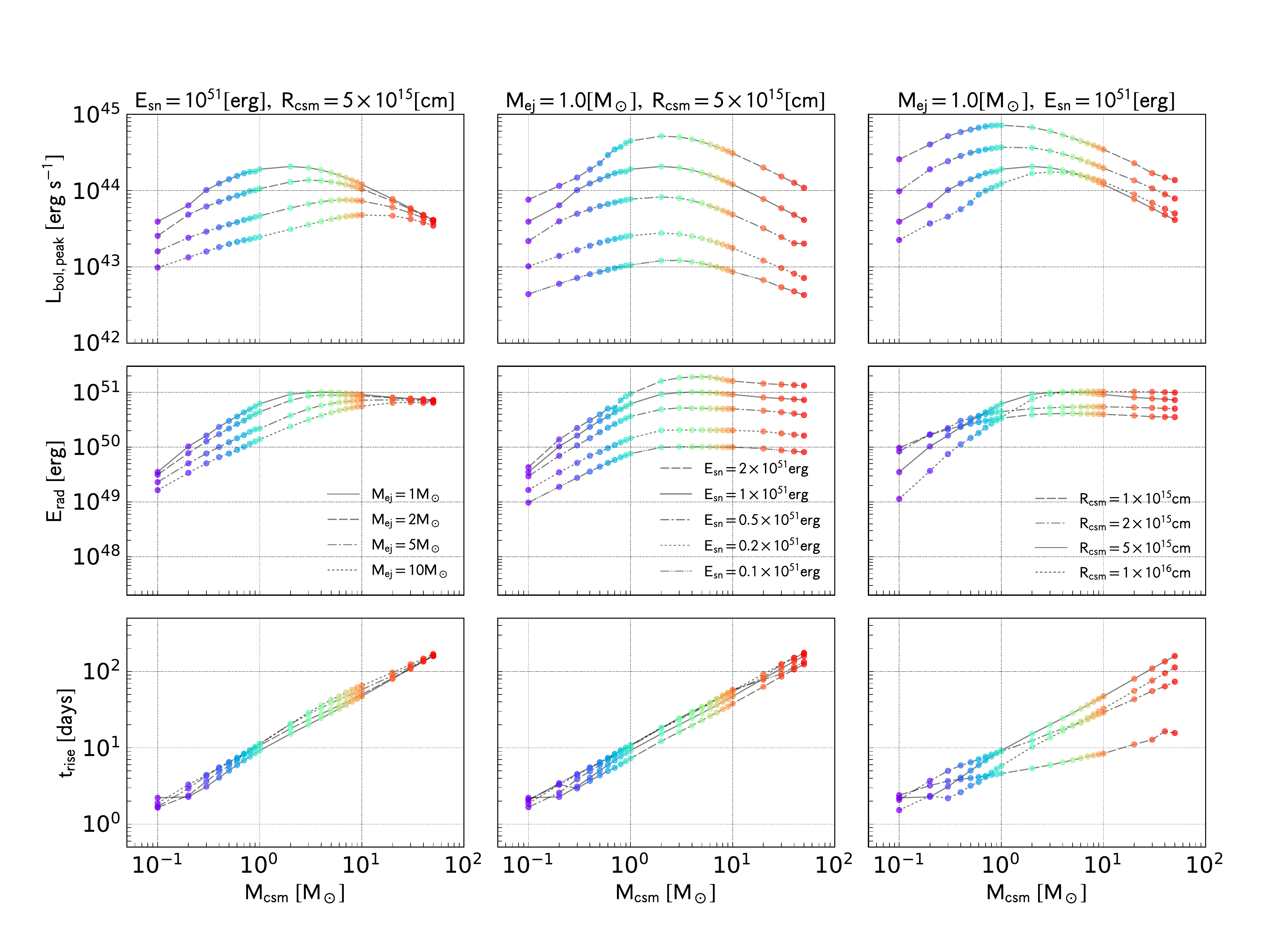}
\caption{Peak bolometric luminosity $L_\mathrm{bol,peak}$ (top), the radiated energy $E_\mathrm{rad}$ (middle), and the rise time $t_\mathrm{rise}$ (bottom) as a function of the CSM mass $M_\mathrm{csm}$. 
The left, center, and right panels represent the dependence on the ejecta mass, the explosion energy, and the outer CSM radius. 
The date points are color-coded in the same way as Figure \ref{fig:lc}. }
\label{fig:scaling_mcsm}
\end{center}
\end{figure*}

One of the important trends in the light curves shown in Figure \ref{fig:lc} and \ref{fig:lc_r1e15} is that the evolutionary timescales of the light curves become longer for larger CSM masses. 
On the other hand, the peak luminosity behaves in a different way. 
For smaller CSM masses, the peak luminosity increases with the CSM mass. 
For larger CSM masses, on the other hand, the peak luminosity gradually decreases with increasing CSM masses. 

In Figure \ref{fig:scaling_mcsm}, we plot the peak bolometric luminosity, the radiated energy, and the rise time as a function of the CSM mass $M_\mathrm{csm}$ for 11 hydrogen-rich model series (253 models in total). 
As shown in the bottom panels, the rise time increases with $M_\mathrm{csm}$. 
This is simply because of the prolonged photon diffusion timescale in a massive and dense CSM. 
The total radiated energy in the middle panel reflects the energy dissipated while the forward shock is still below the photosphere. 
It increases with $M_\mathrm{csm}$ up to a characteristic mass and then approaches a constant value. 
The behavior of the peak bolometric luminosity in the top panels follows those of the rise time and the total radiated energy. 
The peak luminosity increases with $M_\mathrm{csm}$ while both rise time and radiated energy increase with $M_\mathrm{csm}$. 
Then it starts declining with $M_\mathrm{csm}$, when the total radiated energy reaches the saturated value. 

These two regimes are separated by the condition $M_\mathrm{csm}\simeq M_\mathrm{ej}$. 
As seen in the left column of Figure \ref{fig:scaling_mcsm}, the transition mass increases with $M_\mathrm{ej}$ from $1M_\odot$ to $10M_\odot$. 
As we have seen in Section \ref{sec:dynamical_evolution}, the ejecta--CSM system evolves differently depending on whether $M_\mathrm{ej}<M_\mathrm{csm}$ or $M_\mathrm{ej}>M_\mathrm{csm}$. 
For the ejecta mass larger than the CSM mass, $M_\mathrm{ej}>M_\mathrm{csm}$, the interaction-powered emission starts escaping into the surrounding space while the reverse shock is still propagating in the outer part of the ejecta, leaving most ejecta unshocked at the time of the shock breakout. 
As a result, only a small fraction of the ejecta kinetic energy is dissipated and used as thermal emission from the photosphere. 
As the CSM mass increases, the mass of the swept-up ejecta at the breakout increases and hence a larger amount of the ejecta kinetic energy is dissipated.   
On the other hand, a sufficiently massive CSM with $M_\mathrm{csm}>M_\mathrm{ej}$ dissipates most of the ejecta kinetic energy. 
As seen in the middle row of Figure \ref{fig:scaling_mcsm}, the saturated values of the total radiated energy are comparable to the assumed ejecta kinetic energies, which clearly indicates that the most of the ejecta kinetic energy has been used for the interaction-powered emission.

\section{Scaling relations for interaction-powered emission\label{sec:phase_space}}
In this section, we consider the scaling relations for the light curve properties, i.e., the peak bolometric luminosity, the total radiated energy, and the rise time, which are shown in Figure \ref{fig:scaling_mcsm} as a function of the CSM mass. 
As we have demonstrated in the previous section, the dynamical evolution of the shock wave driven by the ejecta can be divided into two regimes, $M_\mathrm{ej}>M_\mathrm{csm}$ and $M_\mathrm{ej}<M_\mathrm{csm}$ (data points in reddish and bluish colors in Figure \ref{fig:scaling_mcsm}). 
We call the former the free-expansion regime, while the latter is called the blast-wave regime. 
After introducing some important concepts in Sections \ref{sec:characteristic_lt} and \ref{sec:efficiency}, we consider these two regimes in Sections \ref{sec:free_expansion_regime} and \ref{sec:blast_wave_regime}, respectively. 

\subsection{Characteristic luminosity and timescale\label{sec:characteristic_lt}}
In the ejecta-CSM collision, the shocked gas is accumulated in the layer between the forward and reverse shock fronts, forming a geometrically thin shell. 
When the shell is in the deep interior of the CSM, photons emitted from the shell experience multiple scattering in the ambient gas, diffusing throughout the CSM toward the photosphere. 
The diffusion velocity of the radiation from the shell at $r=R_\mathrm{s}$ is given by $c/\tau_\mathrm{csm}(R_\mathrm{s})$, where the optical depth is obtained by Equation \ref{eq:tau_csm}. 
Thus, the diffusion timescale at $t$ is estimated as the time required for the radiation front travels from $r=R_\mathrm{s}$ to the photosphere $r=R_\mathrm{ph}$ at the diffusion velocity:
\begin{equation}
    t_\mathrm{diff}(t)
    =\frac{(R_\mathrm{ph}-R_\mathrm{s})\tau_\mathrm{csm}(R_\mathrm{s})}{c},
\end{equation}
where the photospheric radius is given as a function of the CSM radius $R_\mathrm{csm}$ by Equation \ref{eq:R_ph}. 
In simple light curve models for SN explosions, the bolometric luminosity reaches its maximum when the expansion timescale $t$ is equal to the diffusion timescale for photons in the SN ejecta and the peak luminosity is given by the energy production rate at the time (so-called Arnett's rule; \citealt{1982ApJ...253..785A}). 
In this context of the interaction-powered emission, the critical timescale $t_\mathrm{cr}$ is determined so that the dynamical time $t$ is equal to the diffusion time for photons in the medium ahead of the interaction layer:
\begin{equation}
    t_\mathrm{cr}=t_\mathrm{diff}(t_\mathrm{cr}).
    \label{eq:t_cr}
\end{equation}
This critical timescale is expected to give the rising timescale during which the luminosity grows to the peak value. 
We evaluate the energy dissipation rate at $t=t_\mathrm{cr}$ and regard it as an estimate for the peak bolometric luminosity. 

\subsection{Photon production efficiency\label{sec:efficiency}}
The energy dissipation rate at the critical time $t=t_\mathrm{cr}$ turns out to be a good estimate for the peak luminosity as long as the post-shock gas is well thermalized. 
In other words, the post-shock gas heated by the shock passage should produce an enough number of photons to maintain the gas-radiation equilibrium within a timescale shorter than the dynamical time \citep[e.g.,][]{2010ApJ...725..904N,2019ApJ...884...87T}. 
We incorporate the effect of the photon production efficiency by the prescription described below. 
The photon production efficiency is evaluated immediately behind the forward shock front, because the forward shock predominantly contributes to the total energy dissipation rate (see the discussion in Appendix \ref{sec:thin_shell_approximation_chevalier}). 

In our numerical simulations, we consider free-free emission as the only process to create photons, which is appropriate in an almost fully ionized gas. The energy production term in the radiation-hydrodynamic equations (in non-relativistic regime) is written as follows,
\begin{equation}
    \dot{e}_\mathrm{rad}=\rho_\mathrm{sh}\kappa_\mathrm{a}a_\mathrm{r}T_\mathrm{g,sh}^4,
\end{equation}
where the local thermodynamic equilibrium has been assumed. We use the post-shock density $\rho_\mathrm{sh}$ given by
\begin{equation}
    \rho_\mathrm{sh}=\frac{\gamma+1}{\gamma-1}\rho_\mathrm{csm},
\end{equation}
where $\rho_\mathrm{csm}$ is the pre-shock CSM density at $r=R_\mathrm{s}$ and the adiabatic index is assumed to be $\gamma=4/3$ (but, see the following discussion). 
On the other hand, the post-shock internal energy at the forward shock is expressed in terms of the pre-shock CSM density $\rho_\mathrm{csm}$ and the velocity of the shell $V_\mathrm{s}$,
\begin{equation}
    e_\mathrm{int}=\frac{\gamma+1}{2(\gamma-1)}\rho_\mathrm{csm}V_\mathrm{s}^2. 
\end{equation}
Since the dissipated energy is first converted into the post-shock gas internal energy, the gas temperature immediately behind the forward shock front is given by
\begin{equation}
    T_\mathrm{g,sh}=\frac{(\gamma-1)^2}{\gamma+1}\frac{\mu m_\mathrm{u}e_\mathrm{int}}{k_\mathrm{B}\rho_\mathrm{csm}}.
\end{equation}
The maximum radiation energy density that the free-free process can produce within the dynamical timescale $t$ is proportional to $\dot{e}_\mathrm{rad}t$. 
When the post-shock internal energy density is smaller than this radiation energy density, $e_\mathrm{int}<\dot{e}_\mathrm{rad}t$, the gas-radiation equilibrium is achieved within a timescale shorter than the dynamical time $t$ and therefore all the internal energy can be used for the radiation energy budget. 
For $e_\mathrm{int}>\dot{e}_\mathrm{rad}t$, however, the free-free process is not enough efficient to achieve the gas-radiation equilibrium and the fraction of the available radiation energy out of the post-shock internal energy is roughly estimated to be $\dot{e}_\mathrm{rad}t/e_\mathrm{int}$. 
Then, we estimate the efficiency of the photon production in the following way,
\begin{equation}
\epsilon_\mathrm{eff}=\mathrm{min}\left(1,\frac{\dot{e}_\mathrm{rad}t}{\eta e_\mathrm{int}}\right),
\label{eq:eps_eff}
\end{equation}
where $\eta$ is a non-dimensional calibration factor, which is determined later (see, Section \ref{sec:free_expansion_regime}). 

The free-free energy production rate and the efficiency factor introduced here are sensitive to the post-shock density and temperature. 
Since the shock structure is continuously evolving and is modified by radiative cooling effects, predicting the radiative efficiency by the above simplified manner is insufficient. 
In radiative shocks, the effective adiabatic index of the post-shock gas approaches $\gamma=1$ instead of $\gamma=4/3$ for photon gas. 
Therefore, the density jump at the shock front is enhanced, $\rho_\mathrm{sh}>7\rho_\mathrm{csm}$, compared with the adiabatic case with $\gamma=4/3$. 
This difference has a non-negligible impact on the free-free energy production rate because it is proportional to the density squared, $\dot{e}_\mathrm{rad}\propto \rho_\mathrm{sh}^2$. 
Nevertheless, since we expect that the scaling relation of the free-free energy production rate to local physical variables holds correctly, we introduce the calibration factor $\eta$, which is adjusted to obtain the agreement between the semi-analytic formulae and the numerical results. 
In other words, the modifications of the adiabatic index, the jump condition, and so on, have been absorbed in the calibration factor. 

\subsection{Free-expansion regime\label{sec:free_expansion_regime}}
\begin{figure*}
\begin{center}
\includegraphics[scale=0.45]{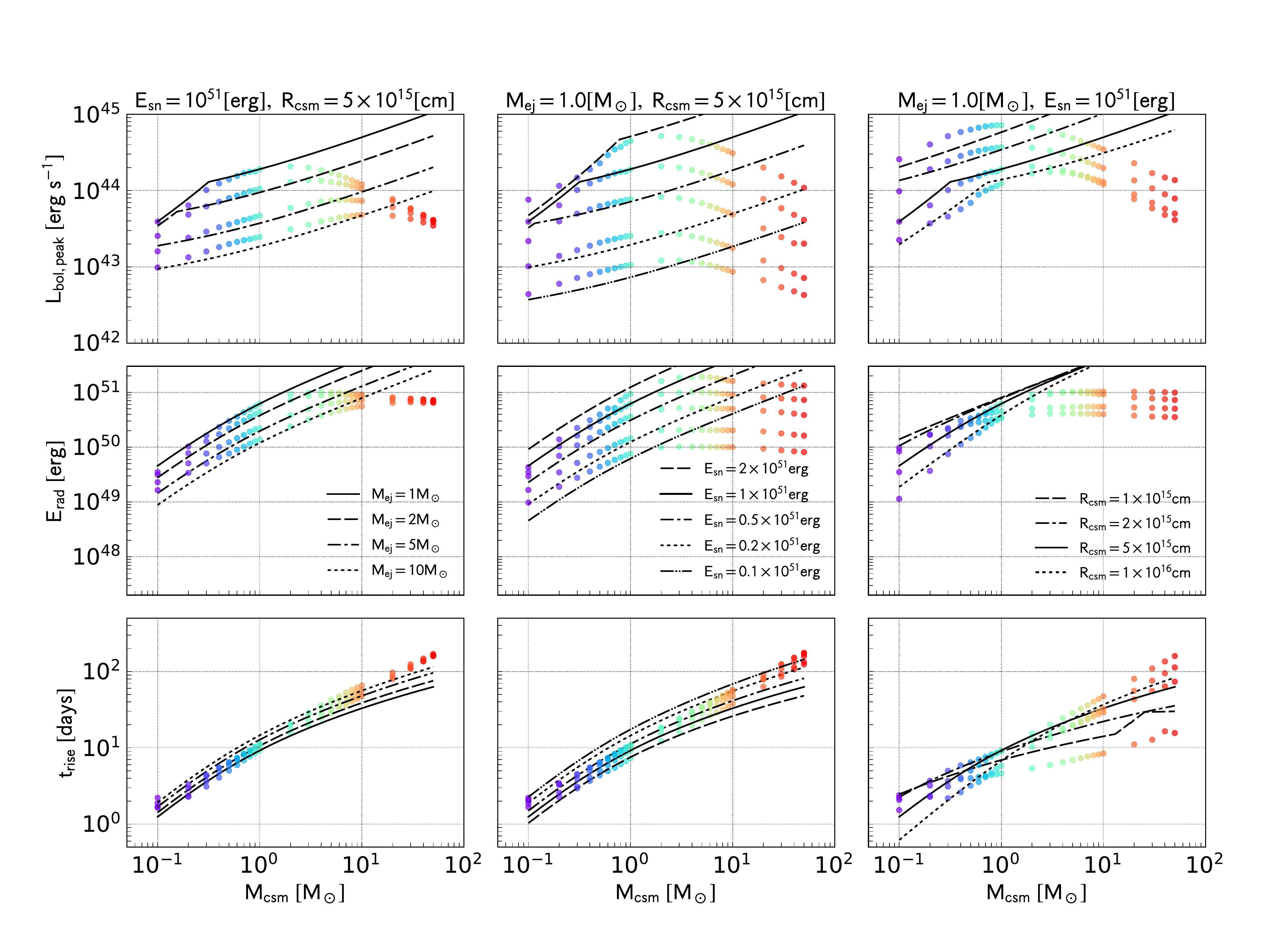}
\caption{Scaling relations for the peak bolometric luminosity, radiated energy, and the rising time. 
In each panel, we compare numerical results (circles) with the semi-analytic scaling relations in the free-expansion regime. 
The date points are color-coded in the same way as Figure \ref{fig:lc}.}
\label{fig:scaling_chevalier}
\end{center}
\end{figure*}

We then consider the case with $M_\mathrm{ej}>M_\mathrm{csm}$ (bluish date points in Figure \ref{fig:scaling_mcsm}).  
In this regime, the most part of the ejecta is still freely expanding, while the outer part of the ejecta is affected by the collision with the CSM. 
The dynamical evolution of the thin shell has been intensively investigated in the literature. 
A series of self-similar solutions are especially useful in describing the expansion of the shell \citep{1982ApJ...258..790C,1982ApJ...259..302C}. 
In Appendix \ref{sec:chevalier}, we summarize some important properties of the self-similar expansion of the shell, which are derived under the so-called thin-shell approximation. 

Using the self-similar expansion law, Equation \ref{eq:rs_ch}, Equation \ref{eq:t_cr} is solved to find the critical time $t_\mathrm{cr}$. 
Then, the rise time is estimated to be 
\begin{equation}
    t_\mathrm{rise,fe}=\epsilon_\mathrm{t,fe}t_\mathrm{cr}. 
\end{equation}
Here we have introduced a non-dimensional factor $\epsilon_\mathrm{t,fe}$ for the overall calibration of the analytic formula. 
Then, we obtain the peak luminosity as follows,
\begin{equation}
    L_\mathrm{peak,fe}=\epsilon_\mathrm{L,fe}\epsilon_\mathrm{eff}\dot{E}_\mathrm{th,fe}(t_\mathrm{cr}),
\end{equation}
where $\dot{E}_\mathrm{th,fe}$ is the internal energy production rate at the shell, Equation \ref{eq:dotE_th}. 
The factor $\epsilon_\mathrm{eff}$ represents the photon production efficiency introduced in Equation \ref{eq:eps_eff}, while the constant $\epsilon_\mathrm{L,fe}$ is another overall calibration factor. 

We expect that the total radiated energy is proportional to the total amount of the internal energy available until the forward shock reaches the photosphere, above which the gas is not strongly coupled with radiation and thus it cannot produce photons efficiently. 
We denote the time at which the shell reaches the photosphere by $t_\mathrm{ph}$,
\begin{equation}
    R_\mathrm{s}(t_\mathrm{ph})=R_\mathrm{ph},
\end{equation}
and then we obtain the total radiated energy as follows,
\begin{equation}
    E_\mathrm{rad,fe}=\epsilon_\mathrm{E,fe}E_\mathrm{th,fe}(t_\mathrm{ph}),
\end{equation}
where $E_\mathrm{th}(t)$ is the internal energy of the shocked gas at $t$ and given by Equation \ref{eq:E_thfe}. 
We again have introduced a non-dimensional factor $\epsilon_\mathrm{E,fe}$. 

We set the numerical factors to be $(\epsilon_\mathrm{t,fe},\epsilon_\mathrm{L,fe},\epsilon_\mathrm{E,fe})=(0.882,0.499,0.415)$. 
The calibration factor for the photon production efficiency is set to $\eta=0.121$. 
We consider the model with $M_\mathrm{csm}=1M_\odot$ in the model series \verb|M1E1R5| as our fiducial model. 
We determine the calibration factors $\epsilon_\mathrm{t,fe}$, $\epsilon_\mathrm{L,fe}$, and $\epsilon_\mathrm{E,fe}$ so that the rise time, the peak luminosity and the total radiated energy of this fiducial model are reproduced. 
This model assumes a sufficiently high CSM density and therefore photons are efficiently produced by free-free emission. 
Therefore, the calibration factors can be determined without the uncertainty associated with the photon production efficiency (i.e., $\epsilon_\mathrm{eff}=1$). 
Models with smaller $M_\mathrm{csm}$ in the same model series suffer from the inefficient photon production. 
Therefore, we use the peak luminosity of the model with $M_\mathrm{csm}=0.1M_\odot$ in the model series \verb|M1E1R5| to determine the calibration factor for the photon production efficiency, $\eta$.

Figure \ref{fig:scaling_chevalier} shows the peak luminosity, the radiated energy, and the rise time estimated by the method described above for different sets of model parameters. 
In this regime, a larger CSM mass produces an increasing amount of the shocked outer ejecta, in which a larger fraction of the kinetic energy is dissipated. 
Therefore, the radiated energy monotonically increases with $M_\mathrm{csm}$. 
The rise time also shows an increasing trend because of a longer diffusion timescale for a larger $M_\mathrm{csm}$. 
The combination of the two trends leads to an increasing $L_\mathrm{peak}$ with $M_\mathrm{csm}$. 
The numerical results in the free-expansion regime (bluish points in each panel) are well reproduced by the semi-analytic scaling relations. 
For larger $M_\mathrm{csm}$ (reddish points), the semi-analytic relation overestimates the peak luminosity and the radiated energy because the assumption of the reverse shock still propagating in the outer ejecta is no longer valid. 
This results in unphysically large radiated energies exceeding the total explosion energy. 

As seen in several $L_\mathrm{peak}$--$M_\mathrm{csm}$ relations shown in the top panels of of Figure \ref{fig:scaling_chevalier}, the slope of the relation becomes steep at small $M_\mathrm{csm}$ and $M_\mathrm{ej}$ in some cases, resulting in a break in the relation. 
This change in the slope is a result of the inefficient photon production in the dilute medium. 
For smaller ejecta and CSM masses, the forward shock propagates in a relatively dilute gas, in which the density is not high enough to produce a sufficient number of photons and to achieve the equilibrium radiation energy density (Section \ref{sec:efficiency}). 
It is remarkable that the slope of the semi-analytic solution in this regime appears to reproduce the trend in the corresponding simulation results. 
The inefficient photon production happens in an extended and less massive CSM. 
As seen in the top right panel of Figure \ref{fig:scaling_chevalier}, the model series with the most extended CSM with $R_\mathrm{csm}=10^{16}$ cm shows a break at $M_\mathrm{csm}\simeq 0.7M_\odot$, suggesting inefficient photon production for $M_\mathrm{csm}<0.7M_\odot$. 
On the other hand, for the model series with the most compact CSM with $R_\mathrm{csm}=10^{15}$ cm, the gas-radiation equilibrium appears to be maintained even for $M_\mathrm{csm}=0.1M_\odot$. 
This different behavior certainly reflects the difference in the CSM density at the shock front, which significantly contributes to the free-free emissivity ($\propto\rho_\mathrm{csm}^2$). 
This finding suggests the presence of an upper limit on the peak bolometric luminosity for a given CSM structure, which is specified by the mass $M_\mathrm{csm}$ and the radius $R_\mathrm{csm}$. 
In the case of the inefficient photon production, the peak luminosity for a fixed $M_\mathrm{csm}$ only slowly increases for increasing ejecta energy, making it difficult to explain bright and rapid transients.

\subsection{Blast-wave regime\label{sec:blast_wave_regime}}
\begin{figure*}
\begin{center}
\includegraphics[scale=0.45]{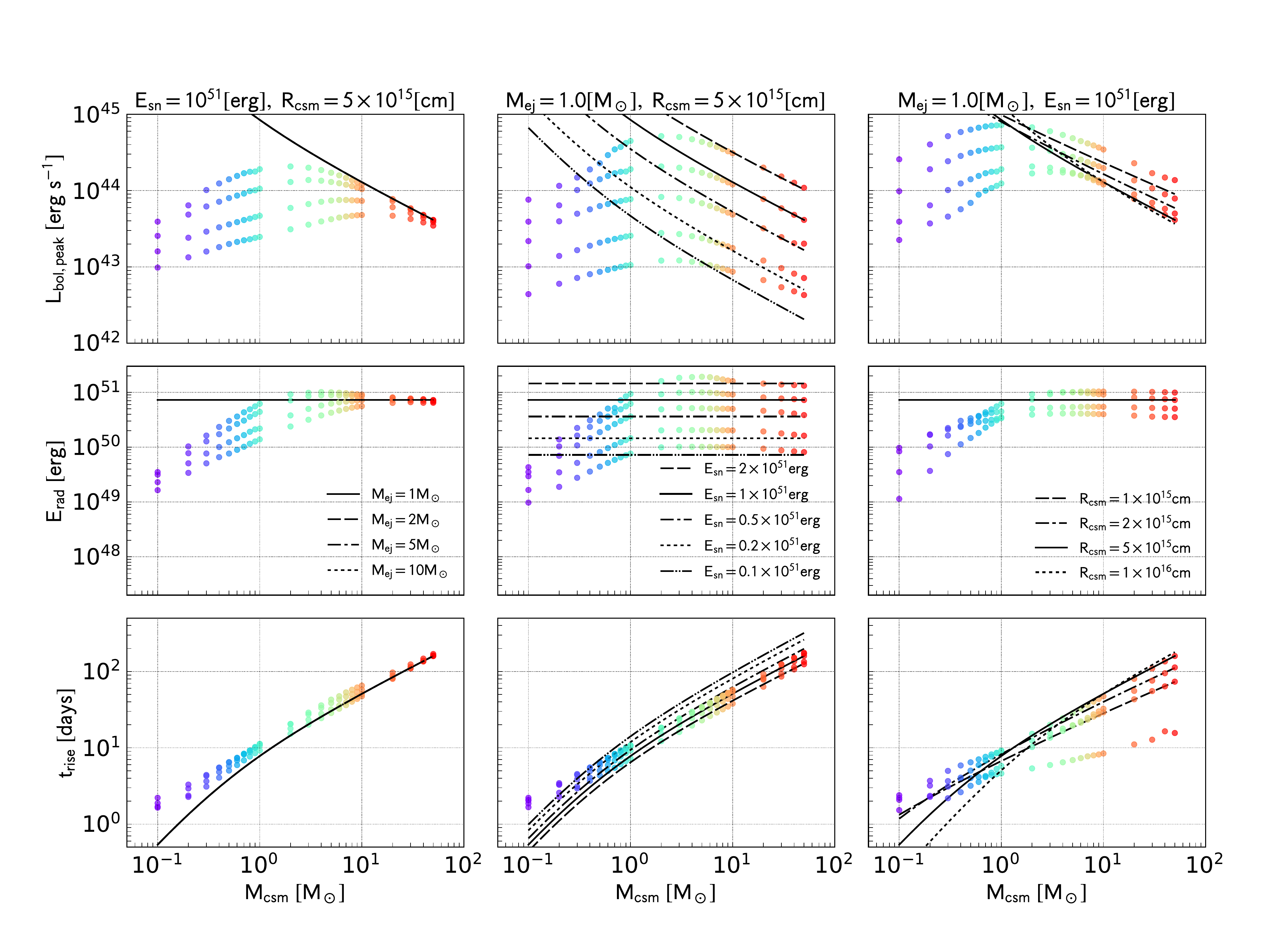}
\caption{Same as Figure \ref{fig:scaling_chevalier}, but for the semi-analytic scaling relations in the blast-wave regime. 
The date points are color-coded in the same way as Figure \ref{fig:lc}.}
\label{fig:scaling_sedov}
\end{center}
\end{figure*}

In this regime, $M_\mathrm{ej}<M_\mathrm{csm}$ (reddish points in Figure \ref{fig:scaling_mcsm}), the situation is similar to a point explosion in a medium with a power-law radial density profile. 
In other words, the mass injected into the CSM is only a small fraction of the CSM mass and the energy of the ejecta is immediately dissipated in a small region at the center of the CSM. 
The forward shock propagation in a power-law atmosphere is well studied in the literature \citep[see, e.g.,][]{1967pswh.book.....Z}. 
In Appendix \ref{sec:sedov}, we again use the thin-shell approximation to briefly derive useful expressions for the forward shock radius, velocity, and so on. 
In this regime, the total radiated energy saturates to a certain fraction of the ejecta kinetic energy. 
Despite the constant radiated energy, the rise time increases with increasing $M_\mathrm{csm}$, which results in decreasing $L_\mathrm{peak}$. 

In a similar way to the free-expansion regime, we use Equation \ref{eq:t_cr} combined with the self-similar expansion law in this regime, Equation \ref{eq:rs_sd}, to find the critical time $t_\mathrm{cr}$. 
Then, we again assume that the rise time is proportional to this critical time, 
\begin{equation}
    t_\mathrm{rise,bw}=\epsilon_\mathrm{t,bw}t_\mathrm{cr}.
\end{equation}
In this blast-wave regime, a constant fraction of the injected energy is converted into the internal energy, Equation \ref{eq:eint_sd}. 
We assume that the total radiated energy is proportional to the internal energy, 
\begin{equation}
    E_\mathrm{rad,bw}=\epsilon_\mathrm{E,bw}E_\mathrm{th,bw},
\end{equation}
where the internal energy $E_\mathrm{th,bw}$ is given by Equation \ref{eq:eint_sd}. 
The peak luminosity is proportional to the internal energy divided by the critical time $t_\mathrm{cr}$:
\begin{equation}
    L_\mathrm{peak,bw}=\epsilon_\mathrm{L,bw}
    \frac{E_\mathrm{th,bw}}{t_\mathrm{cr}}. 
\end{equation}

For the model calibration, we use the model with the largest $M_\mathrm{csm}=50M_\odot$ in the model series \verb|M1E1R5|, which satisfies the condition $M_\mathrm{csm}\gg M_\mathrm{ej}$. 
We simply adjust the three numerical factors $(\epsilon_\mathrm{t,bw},\epsilon_\mathrm{L,bw},\epsilon_\mathrm{E,bw})$ so that the peak luminosity, the radiated energy, and the rise time of the model are reproduced by the scaling relation. 
We obtain $(\epsilon_\mathrm{t,bw},\epsilon_\mathrm{L,bw},\epsilon_\mathrm{E,bw})=(1.15,0.599,0.879)$. 

In Figure \ref{fig:scaling_sedov}, we compare the semi-analytic scaling relations for $L_\mathrm{peak,bw}$, $E_\mathrm{rad,bw}$, and $t_\mathrm{rise,bw}$ with the simulation results. 
For the dependence on the ejecta mass $M_\mathrm{ej}$ (the left column) and the ejecta energy $E_\mathrm{sn}$ (the middle column), the semi-analytic scaling relations well reproduces the simulation results (reddish points in each panel). 
In this regime, the peak luminosity and the radiated energy do not depend on the ejecta mass, which is much smaller than the CSM mass and therefore has a negligible impact on the emission. 
As we have mentioned above, the characteristic quantities converge to certain values in this regime, which are well predicted by the semi-analytic formulae. 
However, some deviation can be found in the dependence on the CSM radius (the right column of Figure \ref{fig:scaling_sedov}). 
In particular, although the semi-analytic scaling relation predicts that the radiated energy does not depend on the CSM radius, the numerical results exhibit some diversity, indicating that the simplified treatment does not reproduce the simulation results perfectly. 

\begin{figure*}
\begin{center}
\includegraphics[scale=0.48]{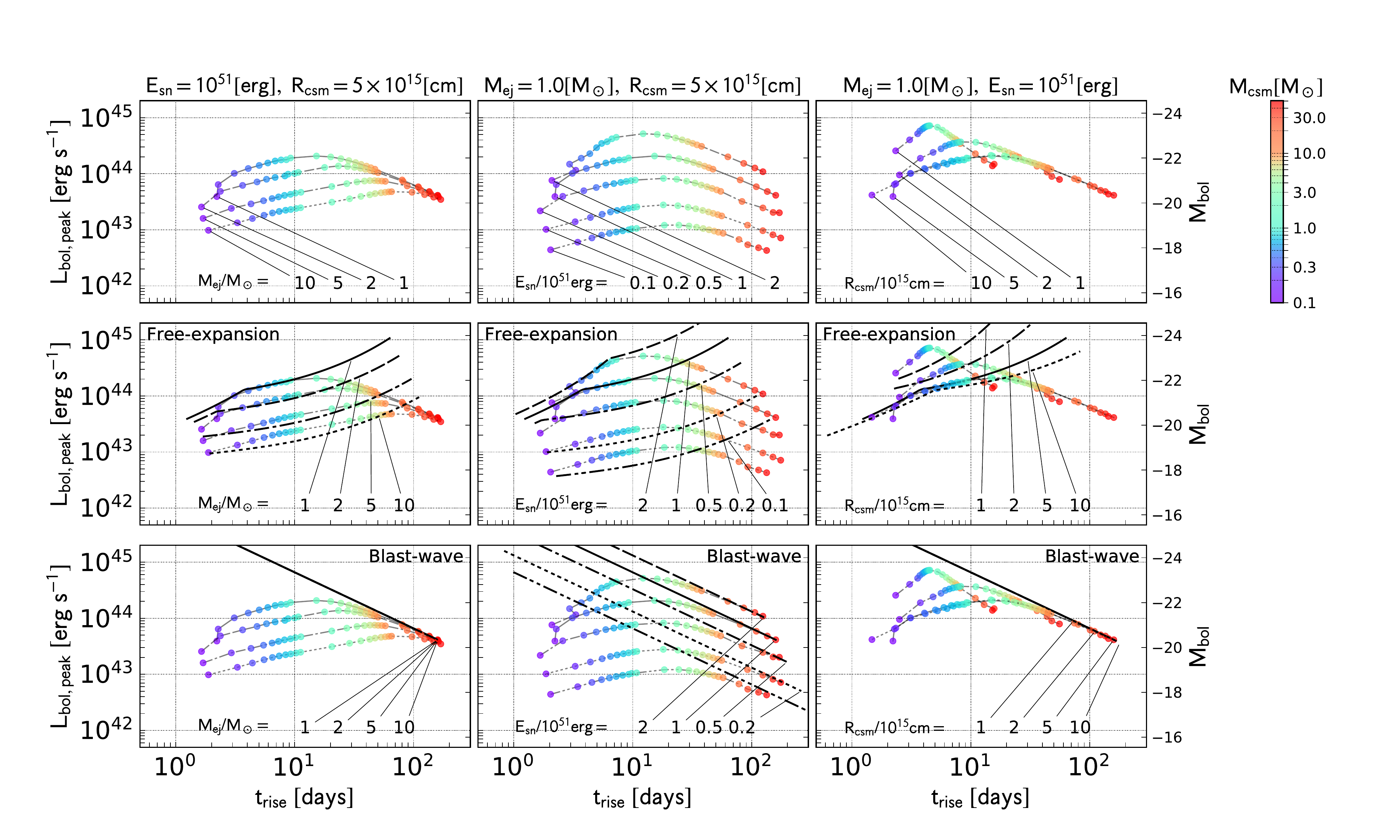}
\caption{Peak bolometric luminosity vs rise time plot. 
In the top row, we present the simulation results with different model parameters. 
The results are compared with semi-analytic $L_\mathrm{peak}$--$t_\mathrm{rise}$ relations in the free-expansion (the middle row) and the blast-wave (the bottom row) regimes. }
\label{fig:Lpeak_Tpeak}
\end{center}
\end{figure*}

\begin{figure*}
\begin{center}
\includegraphics[scale=0.48]{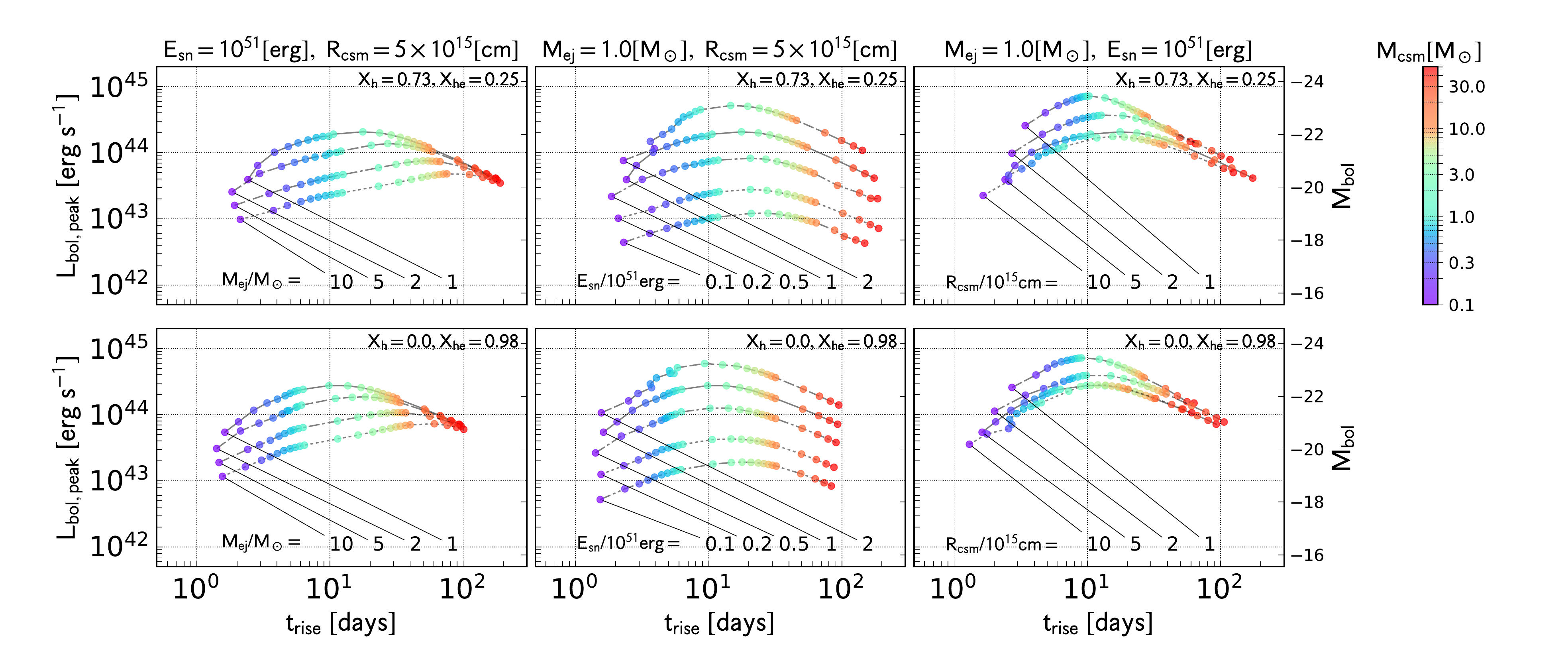}
\caption{Comparison of the $L_\mathrm{peak}$--$t_\mathrm{rise}$ plots for models with hydrogen-rich (upper panels) and hydrogen-free (lower panels) media. }
\label{fig:hydrogen_poor}
\end{center}
\end{figure*}

\begin{figure*}
\begin{center}
\includegraphics[scale=0.48]{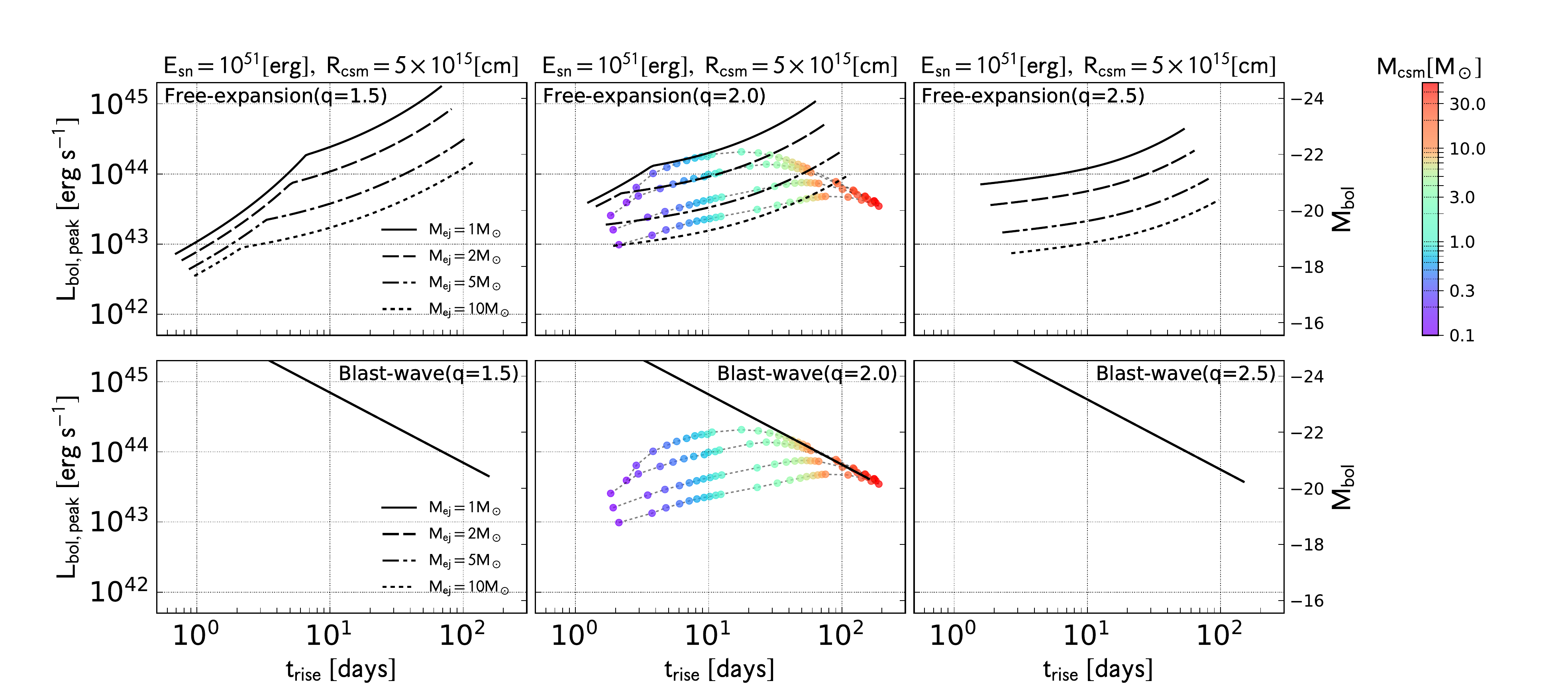}
\caption{Dependence on the slope $q$ of the CSM. 
The analytically obtained $L_\mathrm{peak}$--$t_\mathrm{rise}$ relations for different $q=1.5$ (left), $2.0$ (center), and $2.5$ (right) are compared for the models with fixed $E_\mathrm{sn}=10^{51}$ erg and $R_\mathrm{csm}=5\times 10^{15}$ cm. 
In each column, the upper and lower panels show the analytic relation in the free expansion and blast wave regimes, respectively. 
The numerical results from the model series with the same $E_\mathrm{sn}$ and $R_\mathrm{csm}$ are plotted in the middle panel.
}
\label{fig:csm_slope}
\end{center}
\end{figure*}

\subsection{Remarks}
Despite the model calibration, there are some disagreements between the semi-analytic and numerical results. 
These disagreements are probably owing to several uncertainties in the semi-analytic modeling employed above.  
First of all, we employ Arnett's rule \citep{1982ApJ...253..785A} to obtain the peak luminosity and the peak time. 
However, it is widely known that the peak luminosity predicted by Arnett's rule disagrees with detailed radiative transfer calculations by a factor of a few \citep[e.g.,][]{2015MNRAS.453.2189D}
Secondly, the self-similar expansion laws derived in Sections \ref{sec:chevalier} and \ref{sec:sedov} describe the dynamical evolution of the shell only in the adiabatic case with a constant adiabatic exponent. 
As we have noted above, however, the forward shock becomes radiative as it approaches the photosphere, where the radiative loss has a significant impact on its expansion. 
Furthermore, all the photons leaving the interaction layer are not observed. 
These disagreements have been partially resolved by introducing some calibration factors. 
In order to explain why the adopted values of these calibration factors can reproduce numerical results, we probably have to take into account the radiative effects mentioned above. 
Nevertheless the semi-analytic scaling relations are useful in understanding the overall trend of the numerical results and how the light curve properties are determined for a given set of the model parameters.  
The comparisons in Figures \ref{fig:scaling_chevalier} and \ref{fig:scaling_sedov} demonstrate that the semi-analytic model works well at least as order-of-magnitude estimations for the peak luminosity, the radiated energy, and the rise time.

\section{Rise time vs peak bolometric luminosity\label{sec:t_rise_vs_L_peak}}
\begin{figure*}
\begin{center}
\includegraphics[scale=0.48]{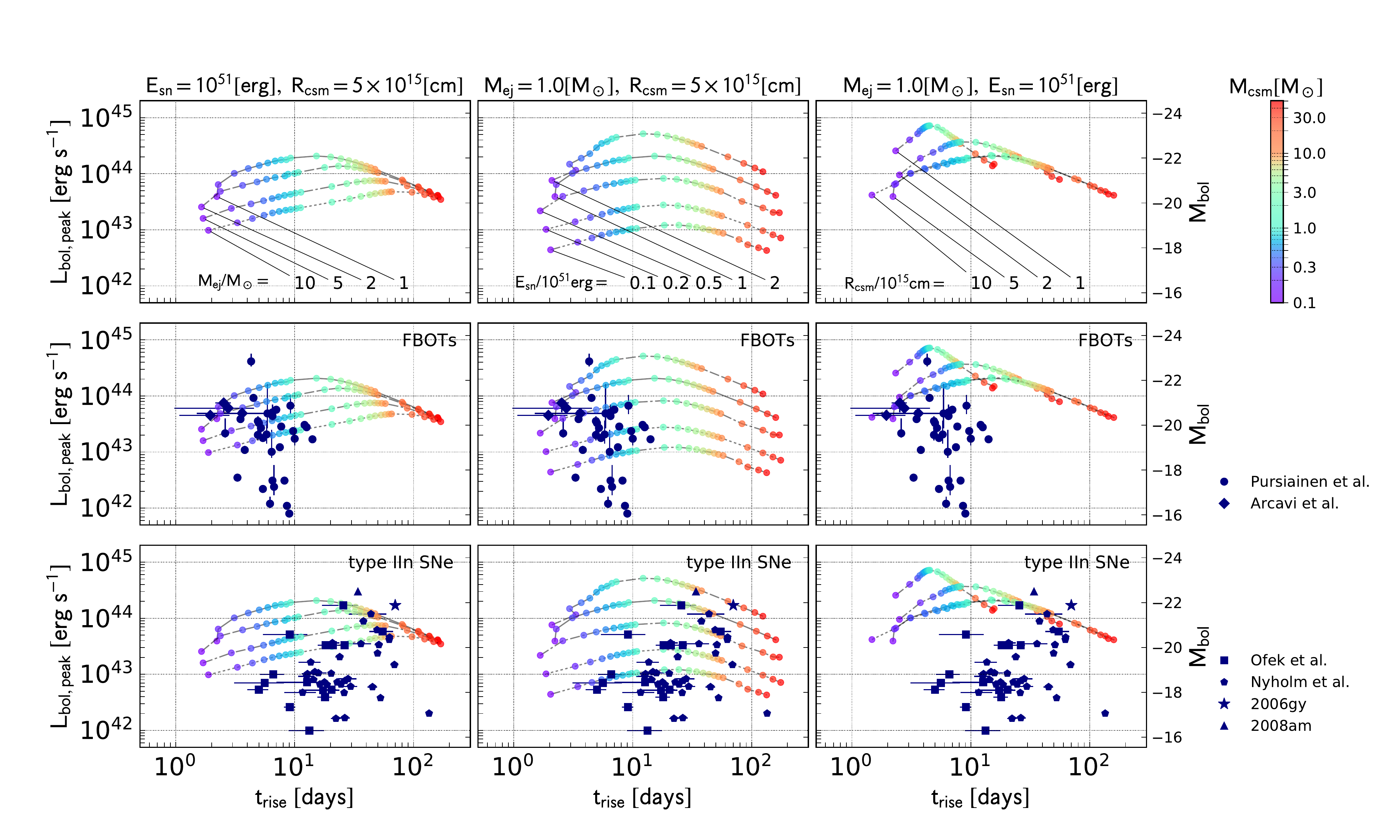}
\caption{Same as Figure \ref{fig:Lpeak_Tpeak}, but we compare the numerical results with observations of FBOTs (middle panels) and type IIn SNe (bottom panels). 
For FBOTs in the middle panels, we plot datasets compiled by \cite{2016ApJ...819...35A} and \cite{2018MNRAS.481..894P}. 
Type IIn SN samples compiled by \cite{2014ApJ...788..154O} and \cite{2019arXiv190605812N} are used in the bottom panels. 
In the bottom panels, two superluminous type IIn SNe, SN 2006gy and 2008am are also plotted. }
\label{fig:Lpeak_Tpeak_obs}
\end{center}
\end{figure*}

\subsection{General trends}
In Figure \ref{fig:Lpeak_Tpeak}, we plot the peak bolometric luminosity as a function of the rise time. 
The peak bolometric luminosity and the rise time span a wide range depending on the model parameters. 
We again emphasize that the peak luminosity shows an increasing trend for short rise times and then a declining trend for long rise times. 
The simulation results are compared with the semi-analytic scaling relations in the free-expansion (middle row) and blast-wave (bottom row) regimes. 
The semi-analytic estimates show good agreements with the simulation results. 

An intriguing point is that the peak luminosities for a fixed ejecta energy $E_\mathrm{sn}$ converge to a single branch for longer rise times (see reddish points in the left and right columns of Figure \ref{fig:Lpeak_Tpeak}). 
The semi-analytic scaling relations in the blast-wave regime (bottom row) well reproduce the branch on which models with a fixed ejecta energy converge, setting an upper bound on the peak luminosity for a given rise time. 
This is again naturally expected in the blast-wave regime. 
As long as the CSM mass is much larger than the ejecta mass, the total radiated energy is simply given by a constant fraction of the injected energy. 
On the other hand, the evolutionary timescale of the light curve is determined by the photon diffusion time in the CSM almost irrespective of the ejecta mass. 
Therefore, the peak luminosity is inversely proportional to the rise time. 
In order to shift this blast-wave limiting luminosity, one has to change the ejecta energy as seen in the middle bottom panel of Figure \ref{fig:Lpeak_Tpeak}, which indicates that the variation in the peak luminosity in this regime can only be produced by the variation in the ejecta energy. 

On the other hand, for short rise times, different sets of the model parameters can produce interaction-powered emission with a wide variety of peak luminosities. 
Particularly, in the range of $L_\mathrm{peak}=10^{43}$--$10^{44}$ erg s$^{-1}$ and $t_\mathrm{peak}=2$--$10$ days, a single set of the peak luminosity and rise time can be explained by multiple models with different model parameters. 
This degeneracy implies that additional information, e.g., the color, the photospheric velocity, and so on, is required to pin-down the appropriate model parameter set reproducing the emission property. 

\subsection{Effects of hydrogen-free media}
We briefly mention results for hydrogen-free media. 
In particular, Fast Blue Optical Transients (FBOTs) may originate from hydrogen-poor or hydrogen-free stellar explosions. 
Without hydrogen, $X_\mathrm{h}=0$, the electron scattering opacity is reduced to $\kappa=0.2$ cm$^2$ g$^{-1}$, which potentially changes the peak luminosity and the evolutionary timescale. 
We also assume a higher recombination temperature, $T_\mathrm{rec}=1.2\times 10^4$ K, for hydrogen-free ejecta. 
We investigate the effects of these modifications on the $L_\mathrm{peak}$--$t_\mathrm{rise}$ relations. 
In Figure \ref{fig:hydrogen_poor}, we compare the simulation results with hydrogen-rich and hydrogen-free media. 
The reduced electron scattering opacity leads to a higher photon diffusion velocity in the CSM. 
As a result, the emission becomes more luminous and short-lived and thus the simulation results cover the upper-left region in the $L_\mathrm{peak}$--$t_\mathrm{rise}$ plot.  
Nevertheless, the differences in the rise time and the peak bolometric luminosity between hydrogen-rich and hydrogen-free models are within a factor of a few. 
Therefore, the general trends discussed in the previous section remain unchanged for hydrogen-free media with reduced electron scattering opacity. 
Considering the limited impact of the different chemical abundance, we again focus on simulation results with hydrogen-rich media in the following. 

\subsection{Effects of CSM density slope}
We also examine how the CSM density slope affects the $L_\mathrm{peak}$--$t_\mathrm{rise}$ relation by using the semi-analytic scaling relations. 
The semi-analytic scaling relations can be applied for wind-like CSMs with the density slope of $1<q<3$ (see, Appendix \ref{sec:self_similar}). 
In Figure \ref{fig:csm_slope}, the scaling relations in the free expansion and blast wave regimes for different CSM density slopes $q=1.5$, $2.0$, and $2.5$ are compared. 
The peak luminosity and the rise time are certainly affected by the different CSM structure. 
In the free expansion regime, the difference in the $L_\mathrm{peak}$--$t_\mathrm{rise}$ relations for $q=1.5$, $2.0$, and $2.5$ is within a factor of a few except for smaller CSM masses. 
For smaller CSM masses, CSMs with shallower density slopes more significantly suffer from the inefficient photon production because of a smaller density in the inner part. 
On the other hand, in the blase wave regime, the $L_\mathrm{peak}$--$t_\mathrm{rise}$ relations are almost independent on the slope $q$.

\subsection{Implications from observations}
In Figure \ref{fig:Lpeak_Tpeak_obs}, we compare the simulation results with currently available samples of FBOTs and type IIn SNe. 
We note that some of the samples are based on single-band observations with a simple bolometric correction. 
Furthermore, the measurements and the definitions of the rise times are based on different methods, which introduces systematic offsets from one sample set to another. 
Although this comparison results should be taken with caution, it offers us possible observational trends among transients potentially powered by the wind shock breakout. 

\subsubsection{FBOTs}
We focus on the comparison between the simulation results and FBOTs (see the middle row of Figure \ref{fig:Lpeak_Tpeak_obs}). 
FBOTs are clustered in a region with shorter $t_\mathrm{rise}$ because of their selection criteria, i.e., evolutionary timescales within $\sim 10$ days.
Their distribution shows a wide variety in the peak luminosity. 
They are different by more than two orders of magnitude from $L_\mathrm{peak}\simeq 10^{42}$ erg s$^{-1}$ to $5\times 10^{44}$ erg s$^{-1}$. 
While the model grids can successfully cover most FBOTs with $L_\mathrm{peak}>10^{43}$ erg s$^{-1}$, less luminous events are difficult to explain. 
Even the least energetic model series with $E_\mathrm{sn}=10^{50}$ erg does not reproduce a peak luminosity as low as $10^{42}$ erg s$^{-1}$, thereby implying that even more extreme conditions, e.g., $E_\mathrm{sn}\ll 10^{51}$ erg, are required. 
As \cite{2018MNRAS.481..894P} suggest, this wide variety in the peak luminosity may indicate that observed FBOTs are actually composed of multiple populations. 
For example, such an extremely less energetic explosion may be realized in a failed supernova with small mass ejection 
\citep{1980Ap&SS..69..115N,2013ApJ...769..109L,2015MNRAS.451.2656K,2018MNRAS.476.2366F,2020arXiv200506103T}. 

Some FBOTs are highly luminous and short-lived. 
In particular, the brightest event in Figure \ref{fig:Lpeak_Tpeak_obs} reaches the peak luminosity as high as $5\times 10^{44}$ erg s$^{-1}$ while its rise time is only $\sim 4$ days. 
In the parameter sets we investigated, this high luminosity can only be achieved when adopting the smallest CSM outer radius, $R_\mathrm{csm}=10^{15}$ cm. 
For extended CSMs with larger radii, the reduced CSM density makes photon production inefficient and such large radiation energy production within a short timescale is difficult to achieve for the explosion energy of the order of $10^{51}$ erg. 
We note that SNe with explosion energies as high as $10^{52}$ erg, i.e., hypernovae, interacting with a small and extended CSM also potentially explain bright and short-lived FBOTs. 
Therefore, these highly luminous and fast-evolving events likely come from explosive phenomena embedded in compact CSMs and/or with large explosion energies if they are actually powered by the wind shock breakout.

\subsubsection{Type IIn SNe}
In the bottom row of Figure \ref{fig:Lpeak_Tpeak_obs}, we compare the numerical results with type IIn SNe. 
Type IIn SNe also show a wide variety in the peak luminosity from $10^{42}$ erg s$^{-1}$ up to a few $10^{44}$ erg s$^{-1}$. 
An overall trend of type IIn SNe in this $L_\mathrm{peak}$--$t_\mathrm{rise}$ plot is that long-lasting events show higher peak luminosity, although the correlation is weak \citep{2019arXiv190605812N}. 
The simulation results successfully explain luminous type IIn SNe with long rising times, while low-luminosity events are difficult to explain. 

It is important to note that luminous type IIn SNe show rise times longer than $20$--$30$ days. 
This is the region where the blast-wave model applies for the parameter sets adopted in numerical simulations. 
As we have seen in the previous section, the model series with a fixed ejecta energy $E_\mathrm{sn}$ converges to a single branch in this regime (left and right columns in Figure \ref{fig:Lpeak_Tpeak_obs}). 
Among the model parameters explored here, changing the ejecta energy is the only way to produce a dispersion in the peak luminosity in this region. 
In fact, the peak luminosities of the two superluminous SNe 2006gy and 2008am in the bottom panels of Figure \ref{fig:Lpeak_Tpeak_obs} can only be explained by models with the ejecta energy of $E_\mathrm{sn}=2\times 10^{51}$ erg (middle bottom panel). 
Therefore, the dispersion in the peak luminosity distribution of long-lasting type IIn SNe may reflect the variation in the kinetic energy of the SN ejecta in massive CSMs since the ejecta-CSM interaction serves as a ``calorimeter'' of the embedded SN ejecta. 
Next, we consider type IIn SNe with rising times of $10$--$30$ days and peak luminosities lower than $10^{43}$ erg s$^{-1}$. 
Even the models with the lowest ejecta energy of $10^{50}$ erg do not explain this low peak luminosity. 
One of the possibilities is that these are very weak explosions with the explosion energy much smaller than the canonical value of $10^{51}$ erg. 
Such less energetic explosions may include non-terminal explosions of massive stars. 
For example, the most slowly evolving object in \cite{2019arXiv190605812N} samples is PTF11qnf with the bolometric peak luminosity of $\sim 2\times 10^{42}$ erg s$^{-1}$ and the rise time longer than $100$ days. 
The underluminous nature of this object implies that the ejecta kinetic energy is extremely small if this is truly a type IIn SN solely powered by the ejecta-CSM interaction. 
As \cite{2019arXiv190605812N} pointed out, however, this object was likely an SN impostor \citep{2000PASP..112.1532V}, i.e., a luminous blue variable(LBV)-like eruption rather than the terminal explosion of a massive star. 
Therefore, the evolutionary timescale of this object may be governed by that of the continuous mass-loss activity rather than the photon diffusion time in the CSM. 
This may imply that sample selection should be done carefully for using them to infer statistical properties of interaction-powered transient populations by using the $L_\mathrm{peak}$--$t_\mathrm{rise}$ plot. 

An alternative possibility is that these type IIn SNe certainly exhibit interaction features but the CSM interaction is not the dominant energy source. 
A peak bolometric luminosity of the order of $10^{42}$ erg s$^{-1}$ and a rise time of $\sim 20$ days can be achieved in normal stripped-envelope SNe \citep{2015A&A...574A..60T,2018A&A...609A.136T,2016MNRAS.458.2973P}, in which the evolutionary timescale is governed by the photon diffusion time in the ejecta and the emission is powered by the nickel radioactive decay. 
In other words, type IIn SNe in any observed sample may include nickel-powered events with hydrogen-rich CSM that is enough massive to produce interaction feature but can not be the dominant power source. 
Such events should show the evolutionary timescale and the peak luminosity similar to normal stripped-envelope SNe or SNe with small hydrogen-rich envelope. 
On the other hand, there are type IIn SNe with high luminosities and long rise times that are less likely explained by nickel-powered emission. 
For such events, CSM interaction clearly plays a critical role in producing their luminous thermal emission and thus we can safely compare our models with them. 
Interestingly, in the analysis by \cite{2019arXiv190605812N}, they suggest that the rise time distribution of their type IIn SN sample can be divided into two populations, fast and slow risers, with the median rise times of $\sim 20$ and $\sim 50$ days. 
This may indicate that the observed type IIn SN sample includes objects predominantly powered by CSM interaction and those showing interaction feature but powered by other energy source(s). 
However, the analysis still suffers from small number statistics and therefore more data should be accumulated. 
When a statistical type IIn SN sample is obtained by on-going or future transient surveys, it is encouraged to test if the sample is composed of multiple populations.

Another important issue is that the trend of type IIn SNe in the $L_\mathrm{peak}$--$t_\mathrm{rise}$ phase space is not simply explained by changing a single parameter, such as the CSM mass and the ejecta energy. 
As seen in Figures \ref{fig:Lpeak_Tpeak}--\ref{fig:Lpeak_Tpeak_obs}, increasing the CSM mass with fixed ejecta mass and energy leads to an anti-correlation between $L_\mathrm{peak}$ and $t_\mathrm{rise}$ (reddish points on the right side of each panel). 
Increasing ejecta mass with a fixed CSM mass and the ejecta energy (data points with the same color in the left columns of Figures \ref{fig:Lpeak_Tpeak}--\ref{fig:Lpeak_Tpeak_obs}) also leads to  anti-correlations with different slopes. 
Furthermore, increasing the ejeta energy with a fixed CSM mass and the ejecta mass (data points with the same color in the middle columns of Figures \ref{fig:Lpeak_Tpeak}--\ref{fig:Lpeak_Tpeak_obs}) leads to an almost vertical line in the plots. 
In other words, these trends obtained by changing a single parameter are not in agreement with the observed trend among $L_\mathrm{peak}$ and $t_\mathrm{rise}$. 
If the positive correlation between $L_\mathrm{peak}$ and $t_\mathrm{rise}$ will be more strongly confirmed in future studies, this disagreement probably indicates that the free parameters considered here, $M_\mathrm{csm}$, $M_\mathrm{ej}$, $E_\mathrm{sn}$, and $R_\mathrm{csm}$ cannot be changed freely, but are somehow related with each other. 
Such underlying relations between the free parameters make a reasonable sense. 
For example, the sum of the ejecta and CSM masses should be related with the initial stellar mass, since producing a massive CSM significantly reduces the ejecta mass. 
Also, observations of CCSNe show a correlation between the ejecta mass and the explosion energy \citep[e.g.,][]{2003IAUS..212..395N}. 
As we have mentioned above, the dispersion in the peak luminosity is predominantly produced by that of the ejecta energy. 
Considering that the rise time is prolonged by increasing the CSM mass, the observed trend may indicate a correlation between the CSM mass and the ejecta energy. 
The CSM outer radius can also change the rise time and thus possibly plays a role in producing the observed trend and/or dispersion. 

Finally, we mention the recent claim for SN 2006gy by \cite{2020Sci...367..415J}. 
They reanalyzed the photometric and spectroscopic data of SN 2006gy and pointed out that SN 2006gy was possibly a type Ia SN interacting with a $\sim10M_\odot$ hydrogen-rich CSM. 
Their recalibration of the bolometric correction suggests that SN 2006gy was less luminous than reported in the literature, reducing the total radiated energy to $9\times 10^{50}$ erg.  
If some luminous type IIn SNe are really type Ia SNe with massive hydrogen-rich CSMs, the blast-wave regime always applies because of the small ejecta mass ($\sim 1M_\odot$) of a type Ia SN compared with that of the surrounding gas. 
Combined with the uniform nature of type Ia SNe, we expect that the peak luminosity and the rise time of such transients show a tight anti-correlation, $L_\mathrm{peak}\propto t_\mathrm{rise}^{-1}$ as we have clarified in this study. 
We note that about $0.6$--$0.7M_\odot$ radioactive nickel is typically synthesized in SNe Ia and deposits as large as $(1.1$--$1.3)\times 10^{50}$ erg of the gamma-ray energy in the ejecta \citep[e.g.,][]{1994ApJS...92..527N}. 
Therefore, the nickel heating could enhance the peak luminosity and the total radiated energy by a few $10\%$ and thus the $L_\mathrm{peak}\propto t_\mathrm{rise}^{-1}$ relation could be elevated in comparison with the pure CSM heating case investigated here.

\section{Summary\label{sec:summary}}
In this work, we have investigated how transients powered by the wind shock breakout occupy specific regions in the phase space of the light-curve properties. 
We particularly focus on the relation between the peak bolometric luminosity and the rise time. 
In order to understand the $L_\mathrm{peak}$--$t_\mathrm{rise}$ relation, we have conducted 1D radiation-hydrodynamic simulations for more than 500 models with various parameter sets specifying the properties of the SN ejecta and the CSM structure. 
Assisted with the semi-analytic scaling relations, we find that the behavior of interaction-powered transients in the $L_\mathrm{peak}$--$t_\mathrm{rise}$ phase space can be divided into free-expansion and blast-wave regimes, depending on the ejecta and CSM masses. 

When the CSM mass is much smaller than the ejecta mass, only the outer ejecta are swept by the reverse shock and the most ejecta remain expanding almost freely. 
In this case, the larger the CSM mass is, the larger energy is dissipated. 
The photon diffusion timescale in the CSM also becomes longer for larger CSM mass. 
This results in increasing trends for the peak bolometric luminosity, the radiated energy, and the rise time for increasing CSM masses. 
In the opposite limit of the CSM mass much larger than the ejecta mass, the energy injected as the kinetic energy of the ejecta is immediately dissipated in the inner part of the CSM. 
The subsequent evolution of the blast wave is well described by a point-explosion solution. 
While the dissipated energy is almost independent on the CSM mass, the photon diffusion timescale increases for increasing CSM mass. 
In this case, the radiated energy is constant, the rise time increases, and the peak bolometric luminosity decreases for increasing CSM mass. 

The $L_\mathrm{peak}$--$t_\mathrm{rise}$ relations obtained by our simulations well cover luminous FBOTs and long-lasting type IIn SNe, although FBOTs with low peak luminosities and type IIn SNe with moderate rise times of $10$--$20$ days are not explained within the adopted parameter sets. 
For FBOTs, the wide variety in the peak luminosity possibly reflects the diversity of their origin. 
Type IIn SNe with the peak luminosity lower than $10^{43}$ erg s$^{-1}$ may be explained by weak SN explosion embedded in a dense CSM. 
An alternative possibility is that they are not solely powered by CSM interaction, but by nickel radioactive decay.

Our analysis and comparisons of the semi-analytic and numerical results demonstrate that the $L_\mathrm{peak}$--$t_\mathrm{rise}$ relation is a promising way to unveil hidden relations among properties of the explosion, such as the explosion energy, the ejecta mass, the CSM mass and radius, and so on. 
Although the available samples of FBOTs and type IIn SNe are still limited, on-going and future transient survey missions will increase the sample size and answer if there are statistically significant correlations among some characteristic quantities of their emission. 
We note, however, that the identification of interacting SNe ultimately relies on spectroscopic follow-up observations. 
Considering the relatively featureless light curves of interacting transients and how they are overlapped with normal and superluminous SNe of other types in the $L_\mathrm{peak}$--$t_\mathrm{rise}$ plot, it would be still challenging to distinguish them from other optical transients with similar evolutionary timescales and luminosities base only on photometric data. 
Therefore, we still need rapid classification procedure based on spectroscopy, although machine learning-based transient classification \citep[e.g.,][]{2016ApJS..225...31L} that combines observed samples of interacting SNe and theoretical models, may help resolving this issue. 
Nevertheless, in the coming LSST era, theoretical models for interaction-powered emission and large observed samples will ultimately help pin-down the origin of interaction-powered transients.

\acknowledgements
We appreciate the anonymous referee for his/her constructive comments on the manuscript. 
A.S. acknowledges support by Japan Society for the Promotion of Science (JSPS) KAKENHI Grand Number JP19K14770. 
This study was also supported in part by the Grants-in-Aid for the Scientific Research of Japan Society for the Promotion of Science (JSPS, Nos. 
JP17H02864, 
JP18K13585, 
JP20H00174, 
JP17H01130, 
JP17K14306, 
JP18H01212  
),  
the Ministry of Education, Science and Culture of Japan (MEXT, Nos. 
JP17H06357, 
JP17H06364 
),
and by JICFuS as a priority issue to be tackled by using Post `K' Computer.
Numerical simulations were carried out by Cray XC50 system operated by Center for Computational Astrophysics, National Astronomical Observatory of Japan. 

\software{Matplotlib (v3.2.1; \citealt{2007CSE.....9...90H})
}

\appendix
\section{Derivation of scaling relations\label{sec:self_similar}}
In this section, we derive the scaling relations for the peak bolometric luminosity and the rise time in both free-expansion and blast-wave regimes. 
For simplicity, we assume a constant opacity in the following. 

\subsection{Circum-stellar medium}
The csm density profile is assumed to be a power-law function of the radius,
\begin{equation}
    \rho_\mathrm{csm}(r)=\frac{(3-q)M_\mathrm{csm}}{4\pi R_\mathrm{csm}^3}
    \left(\frac{r}{R_\mathrm{csm}}\right)^{-q}
    \equiv Dr^{-q},
\end{equation}
for $r<R_\mathrm{csm}$ with $1<q<3$. 
For simplicity, we assume a sharp cut-off in the CSM density at $r=R_\mathrm{csm}$ rather than introducing the exponential factor as in Equation \ref{eq:rho_csm}. 
The opacity $\tau_\mathrm{csm}(r)$ of the CSM at radius $r$ is given by
\begin{equation}
    \tau_\mathrm{csm}(r)=\int^{R_\mathrm{csm}}_r \kappa\rho_\mathrm{csm}(r)dr=
    \frac{\kappa D}{q-1}\left(r^{1-q}-R_\mathrm{csm}^{1-q}\right)
    =\frac{3-q}{4\pi(q-1)}\frac{\kappa M_\mathrm{csm}}{R_\mathrm{csm}^2}
    \left[\left(\frac{r}{R_\mathrm{csm}}\right)^{1-q}-1\right],
    \label{eq:tau_csm}
\end{equation}
where the opacity $\kappa$ is assumed to be a constant. 
The condition $\tau_\mathrm{csm}(r)=1$ gives the photospheric radius $R_\mathrm{ph}$:
\begin{equation}
    R_\mathrm{ph}=
    \left[1+\frac{4\pi(q-1)}{3-q}\frac{R_\mathrm{csm}^2}{\kappa M_\mathrm{csm}}\right]^{-\frac{1}{q-1}}R_\mathrm{csm}.
    \label{eq:R_ph}
\end{equation}

\subsection{Free-expansion regime\label{sec:chevalier}}
The following derivations are based on the thin-shell model for the collision of freely expanding ejecta with a dilute wind medium \citep{1982ApJ...259..302C}. 
Although there exist self-similar solutions discovered by \cite{1982ApJ...258..790C} for this problem, we employ the so-called thin shell approximation to obtain some scaling relations among characteristic physical variables. 
The numerical coefficients for the scaling relations obtained under the approximation slightly differ from those of the exact self-similar solution. 
However, such differences can be absorbed into some numerical factors introduced to calibrate the semi-analytic scaling relations with numerical results.

\subsubsection{Thin shell approximation\label{sec:thin_shell_approximation_chevalier}}
We consider a spherical shell with an infinitesimal thickness between a cold freely expanding ejecta and a spherical CSM at rest. 
In this regime, the shell is located at the outer part of the ejecta with a steep density slope. 
Therefore, we assume the following power-law density profile,
\begin{equation}
    \rho_\mathrm{ej}(r,t)=Br^{-n}t^{n-3},
\end{equation}
with $n\geq 5$.

The radius and velocity of the shell are denoted by $R_\mathrm{s}(t)$ and $V_\mathrm{s}(t)$. 
From the shock jump conditions, the forward shock velocity and the post-shock pressure are expressed in the following way:
\begin{equation}
    V_\mathrm{fs}=\frac{\gamma+1}{2}V_\mathrm{s},
\end{equation}
and
\begin{equation}
    P_\mathrm{fs}=\frac{2}{\gamma+1}\rho_\mathrm{csm}(R_\mathrm{s})V_\mathrm{fs}^2
    =\frac{\gamma+1}{2}\rho_\mathrm{csm}(R_\mathrm{s})V_\mathrm{s}^2.
\label{eq:P_fs}
\end{equation}
In a similar way, the reverse shock velocity and the post-shock pressure are given by
\begin{equation}
    V_\mathrm{rs}=\frac{\gamma+1}{2}V_\mathrm{s}
    -\frac{\gamma-1}{2}\frac{R_\mathrm{s}}{t},
\end{equation}
and
\begin{equation}
    P_\mathrm{rs}=\frac{2}{\gamma+1}\rho_\mathrm{ej}(t,R_\mathrm{s})\left(V_\mathrm{rs}-\frac{R_\mathrm{s}}{t}\right)^2
    =\frac{\gamma+1}{2}\rho_\mathrm{ej}(t,R_\mathrm{s})\left(V_\mathrm{s}-\frac{R_\mathrm{s}}{t}\right)^2.
\end{equation}
When the density profiles of the ejecta and CSM are both power-law function of the radius, the expansion of the thin shell becomes self-similar, $R_\mathrm{s}\propto t^{\alpha}$ and $V_\mathrm{s}=\alpha R_\mathrm{s}/t$. 
The post-shock pressures $P_\mathrm{fs}$ and $P_\mathrm{rs}$ yield
\begin{equation}
    P_\mathrm{fs}=\frac{\gamma+1}{2}\alpha^2Dt^{-2}R_\mathrm{s}^{2-q},
\end{equation}
and
\begin{equation}
    P_\mathrm{rs}=\frac{\gamma+1}{2}(\alpha-1)^2Bt^{n-5}R_\mathrm{s}^{2-n}.
\end{equation}
The self-similarity of the flow requires that the forward and reverse shock pressures, $P_\mathrm{fs}$ and $P_\mathrm{rs}$, evolve with time in the same power-law manner. 
In the thin-shell approximation, we require the pressure balance $P_\mathrm{fs}=P_\mathrm{rs}$, although they are different by a factor of a few in the exact self-similar solution \citep{1982ApJ...258..790C}. 
The pressure balance gives the following exponent,
\begin{equation}
    \alpha=\frac{n-3}{n-q}.
\end{equation}
Thus, one obtains
\begin{equation}
    R_\mathrm{s}=\left(\frac{q-3}{n-3}\right)^{2/(n-q)}\left(\frac{B}{D}\right)^{1/(n-q)}t^{(n-3)/(n-q)},
    \label{eq:rs_ch}
\end{equation}
and
\begin{equation}
    V_\mathrm{s}=\frac{n-3}{n-q}\left(\frac{q-3}{n-3}\right)^{2/(n-q)}\left(\frac{B}{D}\right)^{1/(n-q)}t^{(q-3)/(n-q)}.
    \label{eq:vs_ch}
\end{equation}

When the shell is located at $r=R_\mathrm{s}(t)$ at time $t$, the masses of the ejecta and CSM swept by the shell are found to be,
\begin{equation}
    M_\mathrm{ej,sw}(t)=4\pi \int_{R_\mathrm{s}}^{\infty}\rho_\mathrm{ej}(t,r)r^2dr
    =\frac{Bt^{n-3}}{n-3}R_\mathrm{s}^{3-n}
    =\left(\frac{n-3}{3-q}\right)M_\mathrm{csm}\left(\frac{R_\mathrm{s}}{R_\mathrm{csm}}\right)^{3-q}, 
\end{equation}
and
\begin{equation}
    M_\mathrm{csm,sw}(t)=4\pi \int_0^{R_\mathrm{s}}\rho_\mathrm{csm}(r)r^2dr
    =\frac{4\pi D}{3-q}R_\mathrm{s}^{3-q}=M_\mathrm{csm}\left(\frac{R_\mathrm{s}}{R_\mathrm{csm}}\right)^{3-q}.
\label{eq:M_csm_sw}
\end{equation}
Thus, the total swept-up mass yields
\begin{equation}
    M_\mathrm{sw}(t)=\frac{n-q}{3-q}M_\mathrm{csm}
    \left(\frac{R_\mathrm{s}}{R_\mathrm{csm}}\right)^{3-q}.
\end{equation}
The kinetic energy of the ejecta component swept up by the shell is calculated as follows,
\begin{equation}
   E_\mathrm{sw}(t)
    =2\pi\int_{R_\mathrm{s}}^{\infty}\rho_\mathrm{ej}(t,r)\frac{r^4}{t^2}dr
    =\frac{2\pi B}{n-5}t^{n-5}R_\mathrm{s}^{5-n}
    =\frac{n-q}{2(n-5)}M_\mathrm{sw}(t)V_\mathrm{s}^2.
\end{equation}
On the other hand, the kinetic energy of the shell is given by
\begin{equation}
    E_\mathrm{kin,fe}(t)=\frac{1}{2}M_\mathrm{sw}(t)V_\mathrm{s}^2=\frac{n-5}{n-q}E_\mathrm{sw}(t)
\end{equation}
The energy conservation gives the thermal energy of the gas in the shell, 
\begin{equation}
    E_\mathrm{th,fe}(t)=E_\mathrm{sw}(t)-E_\mathrm{kin}(t)
    =\frac{5-q}{n-q}E_\mathrm{sw}(t)
    \equiv f_\mathrm{th}E_\mathrm{sw}(t),
    \label{eq:E_thfe}
\end{equation}
which is assumed to be dominated by radiation energy and thus used for interaction-powered emission. 
The derivative of the thermal energy with respect to time leads to
\begin{equation}
    \dot{E}_\mathrm{th,fe}(t)=\frac{(q-3)(5-n)}{n-q}
    \frac{E_\mathrm{th,fe}(t)}{t},
    \label{eq:dotE_th}
\end{equation}
which gives the radiation energy increase per unit time.

Finally, we consider the relative contributions of the forward and reverse shocks on the energy dissipation rate. 
The energy dissipation rates at the forward and reverse shock fronts are proportional to $\rho_\mathrm{csm}(R_\mathrm{s})V_\mathrm{s}^3$ and $\rho_\mathrm{ej}(t,R_\mathrm{s})(R_\mathrm{s}/t-V_\mathrm{s})^3$ (excluding the commonly appearing coefficient). 
From the previously assumed pressure balance, the ratio of reverse to forward shock contribution reduces to
\begin{equation}
    \frac{\rho_\mathrm{ej}(t,R_\mathrm{s})(R_\mathrm{s}/t-V_\mathrm{s})^3}
    {\rho_\mathrm{csm}(R_\mathrm{s})V_\mathrm{s}^3}=\frac{R_\mathrm{s}/t-V_\mathrm{s}}{V_\mathrm{s}}=\frac{1-\alpha}{\alpha}=\frac{3-q}{n-q}. 
\end{equation}
For $n=10$ and $q=2$, the ratio leads to $1/8$, which means that the reverse shock only insignificantly contributes to the total energy dissipation rate.

\subsection{Blast-wave regime\label{sec:sedov}}
We consider the case with $M_\mathrm{ej}\ll M_\mathrm{csm}$. 
In this case, the whole ejecta are immediately swept by the reverse shock and thus depositing all the energy in a central small region of the massive CSM, which is converted into the kinetic and internal energies of the shocked gas. 
The consequence is regarded as a point-like explosion in a power-law medium. 
Although there exists a self-similar solution for the problem \citep[see, e.g.,][]{1967pswh.book.....Z}, we again treat the blast wave propagation in an approximate manner. 

\subsubsection{Blast wave propagation}
The blast wave propagation is treated by a thin shell approximation, in which the swept-up gas is concentrated within a spherical shell with a small width. 
The radius and the velocity of the shell are again denoted by $R_\mathrm{s}(t)$ and $V_\mathrm{s}(t)$ and we assume their power-law dependence on $t$, $R_\mathrm{s}(t)=At^\alpha$ and $V_\mathrm{s}(t)=\alpha At^{\alpha-1}$. 
As we have seen in Section \ref{sec:chevalier}, the post-shock pressure of the forward shock is given by Equation \ref{eq:P_fs}. 
Assuming that the pressure distribution in the post-shock region is uniform, which is a good approximation for the causally connected shocked region, the internal energy of the shocked region is obtained as follows,
\begin{equation}
    E_\mathrm{th,bw}=\frac{P_\mathrm{fs}}{\gamma-1}\frac{4\pi}{3}R_\mathrm{s}^3
    =\frac{(3-q)(\gamma+1)}{3(\gamma-1)}M_\mathrm{csm}
    \left(\frac{R_\mathrm{s}}{R_\mathrm{csm}}\right)^{3-q}V_\mathrm{s}^2.
\end{equation}
On the other hand, the mass of the swept-up gas is given by Equation \ref{eq:M_csm_sw}. 
Therefore, the kinetic energy of the swept-up gas leads to
\begin{equation}
    E_\mathrm{kin,bw}=\frac{1}{2}M_\mathrm{csm,sw}V_\mathrm{s}^2=\frac{1}{2}M_\mathrm{csm}
    \left(\frac{R_\mathrm{s}}{R_\mathrm{csm}}\right)^{3-q}V_\mathrm{s}^2. 
\end{equation}
The sum of these two equation should be equal to the total energy $E_\mathrm{sn}$:
\begin{equation}
    E_\mathrm{sn}=\left[\frac{1}{2}+\frac{(3-q)(\gamma+1)}{3(\gamma-1)}\right]M_\mathrm{csm}
    \left(\frac{R_\mathrm{s}}{R_\mathrm{csm}}\right)^{3-q}V_\mathrm{s}^2
\end{equation}
This relation gives the following exponent
\begin{equation}
    \alpha=\frac{2}{5-q},
\end{equation}
and coefficient
\begin{equation}
    A=\left[
    \frac{3(\gamma-1)(5-q)^2}{3+9\gamma-2q-2\gamma q}
    \right]^{\frac{1}{5-q}}
    R_\mathrm{csm}^{\frac{3-q}{5-q}}
    M_\mathrm{csm}^{-\frac{1}{5-q}}
    E_\mathrm{sn}^{\frac{1}{5-q}}.
\end{equation}
Therefore, the forward shock radius and the shock velocity are expressed as follows,
\begin{equation}
    R_\mathrm{s}(t)=\left[
    \frac{3(\gamma-1)(5-q)^2}{3+9\gamma-2q-2\gamma q}
    \right]^{\frac{1}{5-q}}
    R_\mathrm{csm}^{\frac{3-q}{5-q}}
    M_\mathrm{csm}^{-\frac{1}{5-q}}
    E_\mathrm{sn}^{\frac{1}{5-q}}
    t^{\frac{2}{5-q}}
    \label{eq:rs_sd}
\end{equation}
and
\begin{equation}
    V_\mathrm{s}(t)=\frac{2}{5-q}\left[
    \frac{3(\gamma-1)(5-q)^2}{3+9\gamma-2q-2\gamma q}
    \right]^{\frac{1}{5-q}}
    R_\mathrm{csm}^{\frac{3-q}{5-q}}
    M_\mathrm{csm}^{-\frac{1}{5-q}}
    E_\mathrm{sn}^{\frac{1}{5-q}}
    t^{-\frac{3-q}{5-q}}
    \label{eq:vs_sd}
\end{equation}
The ratios of the internal and kinetic energies to the total energy are always constant,
\begin{equation}
    E_\mathrm{th,bw}=\frac{(3-q)(\gamma+1)}{3+9\gamma-2q-2\gamma q}E_\mathrm{sn},
    \label{eq:eint_sd}
\end{equation}
and 
\begin{equation}
    E_\mathrm{kin,bw}=\frac{3(\gamma-1)}{3+9\gamma-2q-2\gamma q}E_\mathrm{sn}.
\end{equation}



\bibliography{refs}



\end{document}